\documentclass[letterpaper,twocolumn,10pt]{article}
\usepackage{usenix-2020-09}

\usepackage{tikz}
\usepackage{amsmath}

\usepackage{filecontents}

\usepackage[normalem]{ulem}
\usepackage{makecell}
\usepackage[ruled,vlined,linesnumbered]{algorithm2e}
\usepackage{pifont}
\usepackage{amsfonts}
\begin{document}
%-------------------------------------------------------------------------------

%don't want date printed
\date{}

\title{\textit{Minions}: Accelerating Large Language Model Inference with Aggregated Speculative Execution}
\author{
{\rm Siqi Wang}
\and
{\rm Hailong Yang\thanks{Corresponding author}}
\and
{\rm Xuezhu Wang}
\and
{\rm Tongxuan Liu}
\and
{\rm Pengbo Wang}
\and
{\rm Xuning Liang}
\and
{\rm Kejie Ma}
\and
{\rm Tianyu Feng}
\and
{\rm Xin You}
\and
{\rm Yongjun Bao}
\and
{\rm Yi Liu}
\and
{\rm Zhongzhi Luan}
\and
{\rm Depei Qian}
} % end author

\maketitle

%-------------------------------------------------------------------------------
\begin{abstract}
Large language models (LLM) have recently attracted surging interest due to their outstanding capabilities across various domains. However, enabling efficient LLM inference is challenging due to its autoregressive decoding that generates tokens only one at a time. Although research works apply pruning or quantization to speed up LLM inference, they typically require fine-tuning the LLM, incurring significant time and economic costs. Meanwhile, speculative decoding has been proposed to use small speculative models (SSMs) to accelerate the inference of LLM. However, the low acceptance rate of SSM and the high verification cost of LLM prohibit further performance improvement of inference. In this paper, we propose \textit{Minions}, an LLM inference system that accelerates LLM inference with a collective and adaptive speculative generation. Specifically, \textit{Minions} proposes a majority-voted mechanism to leverage multiple SSMs to jointly speculate the outputs of LLM, which improves the inference performance without introducing prohibitive computation costs for LLM. To better trade off the number of tokens speculated from SSM and the verification cost of LLM, \textit{Minions} proposes an adaptive mechanism to dynamically determine the optimal speculation length of SSM, which can achieve better inference performance across different models, datasets, and hyper-parameters. In addition, \textit{Minions} decouples the SSM decoding and LLM verification efficiently and adopts a pipelined execution mechanism to further improve the inference performance of LLM. By comparing with the state-of-the-art LLM inference systems, we demonstrate that \textit{Minions} can achieve higher inference throughput and lower inference time.
\end{abstract}

%Since LLM is used as a verifier instead of an incremental decoder, \textit{Minions} can overlap the execution of LLM and small language models, which significantly improves the inference throughput. 

\section{Introduction}
% 大语言模型（LLMs）如GPT，LLama，OPT近年来在产业界和学术界的多个领域取得了巨大成功。这些LLMs的成功应用依赖于其核心在语言文本上具有出色理解能力的transformer架构。然而，LLMs通常需要使用海量数据规模进行训练，并且其模型参数规模超过100B规模，进而在推理过程中也需要惊人的算力。这导致LLM推理需要消耗大量能源，产生巨大的碳排放以及经济开销。例如，研究公司SemiAnalysis表示，OpenAI公司需要3617台英伟达的HGX A100服务器，总共28936个GPU来支持ChatGPT，这意味着每天的能源需求就要达到564兆瓦时~\cite{de2023growing}，导致很大的碳排放量（contributing significantly to carbon footprint）。对于OpenAI最新的GPT-4模型，处理1000个tokens的input和output分别需要＄0.03和＄0.06的推理成本~\cite{openaiprice}，这昂贵的成本让很多中小企业难以负担进而限制了大语言模型领域的进一步应用创新。因此，如何高效且低成本的部署这些LLMs是目前大模型应用所面临的一个巨大的挑战。
Recently, Large Language Models (LLMs) such as GPT~\cite{achiam2023gpt}, Llama~\cite{touvron2023llama}, and OPT~\cite{zhang2022opt} have achieved significant success in various domains within both industry and academia~\cite{adiwardana2020towards, roller2020recipes, nallapati2016abstractive, paulus2017deep, see2017get, chen2021evaluating, xu2015show, yang2016review}. The success of these LLMs relies on their core model architecture of transformer~\cite{vaswani2017attention}, which exhibits unexceptional natural language understanding capabilities. However, LLMs typically require extensive training with massive datasets, and their model parameter can easily scale beyond hundreds of billions (e.g., OPT-175B), demanding substantial computational power during the inference process. The surging computational demands also result in astonishing energy consumption for LLM inference, contributing to significant carbon emissions and economic expenses. For example, research firm SemiAnalysis indicated that OpenAI required 3,617 NVIDIA HGX A100 servers, with a total of 28,936 GPUs, to support the popular inference service of ChatGPT. This also implies a daily energy demand of 564 MWh~\cite{de2023growing}, contributing a large amount of carbon footprint. For the latest OpenAI GPT-4 model, the inference cost for processing 1,000 tokens of input and output is reported to be \$0.03 and \$0.06, respectively~\cite{openaiprice}. The high cost from both environment and economics makes it formidable for many small and medium-sized enterprises to afford, hindering further innovation in the applications of LLMs. Therefore, to support the LLM inference efficiently is vital for the future success of LLMs.

%（备注）（大challenge）自回归→*incremental decoding→串行计算资源利用程度低，一方面大模型执行由于要拿KV本身memory bound 更加剧了这一情况→低效

% 特别是，LLM推理采用自回归式的生成方法，每次只会根据上文预测下一个token。其推理过程分为两个阶段，prefill和decode。在prefill阶段，LLM以prompt(input sequence)作为输入，生成一个token。在decode阶段，LLM以上一次生成的token作为输入，并新生成一个token，该token会被添加到generated tokens序列的末尾。因此对于一条请求，为了生成n个输出，大模型需要顺序执行n个iteration，造成单条请求执行时并行度和计算资源利用率低的问题。与此同时，大模型执行时多条请求的并发度同样受限，加剧了推理的低计算资源利用率问题。具体来说，LLM所采用的attention机制需要获取之前所有tokens的keys和values以计算新生成token的输出。为了避免重复计算，在LLM推理时会使用缓存机制将这些keys和values存储在KVCache中以便后续计算重复使用。KVCache在大模型中的普遍使用导致了两方面的问题：一是导致每个iteration执行过程中都需要加载很大的K和V矩阵，占用内存带宽，使得推理性能受限于内存带宽（memory-bounded）；二是KV Cache占用的显存会随着推理的执行显著增加，推理过程中batch size不得不保持较小的设置（例如，4、8等），进而严重限制了LLM的推理吞吐。
In particular, the token generation of LLM inference is performed iteratively in an autoregressive manner, that decodes the next token based on the preceding token sequence in each iteration. Specifically, the inference process can be divided into two stages: prefill and decode. In the prefill stage, LLM takes the prompt (e.g., input sequence) as input and generates one token. In the decode stage, LLM uses the previously generated token as input to produce a new token, which is then appended to the end of the generated token sequence. The above process is performed iteratively, therefore for a single request to generate $n$ outputs, LLM needs to execute $n$ iterations sequentially, resulting in low parallelism and underutilization of computation resources during the inference. 

Meanwhile, the concurrency of requests during the LLM inference is also quite limited, exacerbating the underutilization of computation resources and thus low inference throughput. The reason can be attributed to the attention mechanism adopted by LLM that requires accessing keys and values for all previous tokens to compute the output of the newly generated token. To avoid redundant computation, a caching mechanism called KVCache is commonly adopted to store previous keys and values for subsequent reuse in future token generation. Despite the elimination of redundant computation, the adoption of KVCache in LLM inference also introduces downsides. On one hand, it needs to load large key and value matrices during each decoding iteration, which consumes memory bandwidth and thus constrains the inference performance due to limited memory bandwidth (a.k.a., memory-bounded). On the other hand, the GPU memory occupied by KVCache increases significantly during inference, preventing the LLM from adopting a large batch size and thus severely restricting the inference throughput.

%（备注）batch size上不去 吞吐量上不去。xxx（总述）。造成了GPU计算资源underutilization并限制了serving的吞吐量。
% （备注）memory bound 计算资源利用不上去 **memory大 计算资源上不去**
% （备注）显存占用特别高 但是计算资源上不去
%（备注）模型压缩 量化 但精度有损且需要一系列的fine-tune 实际生产中对大模型不是一个很经济实用的方案，还有一种方案是spec spec对模型本身不需要进行任何改动，属于经济实际的方案。 过度一下 再引到spec上

% 为了解决大模型推理吞吐量低，无法充分利用计算资源的问题，研究人员提出了多种大模型推理优化方案。一些方法[Medusa]通过在LLM中引入额外的decoding heads从而并行预测多个后续token，从而加速LLM推理过程。但其需要对大模型进行重训练，产生难以忽略的训练开销。另外一些方法[x]观察到并非所有任务都需要full-sized的大模型，可以通过对大模型进行剪枝或量化从而对结果进行近似。尽管这些解决方案可以减少显存的占用，提高推理的batch size进而提升吞吐。但该方法会造成精度损失，限制了其应用场景。此外，这类方法通常需要对大模型进行re-train从而达到所需的精度，这从经济成本上考虑非常不实用，且很少有人能负担得起。speculative decoding作为一种新兴的大模型推理优化方案借助小模型加速大模型的推理过程，且它无需对大模型本身进行任何修改，因此更加经济实用。具体来说，speculative decoding使用一个更小且更快的模型(SSM)生成请求的下k个token(其中k是可调整的超参数，后文称为speculation length)，并使用原始大模型进行并行验证。由于大模型验证的效率远高于顺序生成的效率，并且可以充分利用大模型推理时闲置的GPU算力资源，因此可以有效提升推理吞吐。speculative decoding带来的性能收益取决于大模型验证一次的acceptance rate（correctly verified token length/speculation length），acceptance rate越高，吞吐提升越大。因此，SSM的speculation accuracy对speculative decoding来说至关重要。不幸的是，我们注意到现有的speculative decoding方法的准确性并不是很高，(Unfortunately, we notice that the accuracy of existing speculative decoding methods is not very high.)，例如采用[X]的方法进行speculative decoding时，小模型的acceptance rate在x数据集下只有xx。此外，现有speculative decoding方法将LLM和SSM分开部署在不同GPU上，且LLM和SSM需要顺序执行。即大模型需要等待小模型推理若干步（speculation length）后才能进行一次验证，小模型也需要等待大模型验证后才能进行下一轮推理，这种大小模型在部署时的紧耦合导致GPU上的计算资源仍没有得到充分利用，限制了大模型推理吞吐的进一步提升。

To address the limitation of low inference throughput of LLMs, researchers have proposed various optimization techniques~\cite{medusa,sanh2020movement,jacob2018quantization,frantar2022optq}. For example, Medusa~\cite{medusa} has been proposed to accelerate the LLM inference by introducing additional decoding heads, enabling the parallel prediction of multiple subsequent tokens. However, such a method requires retraining the LLM, incurring prohibitive training costs. Other methods~\cite{sanh2020movement,jacob2018quantization,frantar2022optq} observe that not all tasks require a full-sized LLM and propose to prune or quantize LLM. While these methods can reduce GPU memory usage of LLM, and thus improve inference throughput with increased batch size, the accuracy loss restricts their applicability in certain scenarios. In addition, these methods typically require fine-tuning the LLM to reach the desired accuracy, which is hardly affordable economically for the majority of researchers and practitioners. 

Speculative decoding~\cite{leviathan2023fast,chen2023accelerating,miao2023specinfer}, as an emerging technique to accelerate LLM inference, harnesses a small speculative model (SSM) to expedite the inference process of LLM. Advantageously, it requires no modifications to the LLM itself and thus does not require additional retraining or fine-tuning of the LLM, rendering it more practical economically. Specifically, speculative decoding adopts SSM to generate the next $s$ tokens for a given request (where $s$ is a tunable hyperparameter, referred to as the \textit{speculation length}) and utilizes the LLM for parallel verification of the generated tokens~\cite{miao2023specinfer}. Since the SSM inference is an order of magnitude faster than the LLM, it can effectively improve the inference latency and throughput if more tokens generated from the SSM can be accepted by the LLM. 

For speculative decoding, the acceptance rate (correctly verified token length/speculation length) is a critical factor impacting the performance of speculative decoding (e.g., a higher acceptance rate means higher inference throughput). Unfortunately, we notice that the acceptance rate of existing speculative decoding methods~\cite{leviathan2023fast,chen2023accelerating,miao2023specinfer} remains low. For example, when evaluating the speculative decoding method proposed by~\cite{miao2023specinfer}, the average acceptance rate of SSMs is only 0.22 (Llama-7B with ChatGPT Prompts~\cite{Chatgpt-prompts}). In addition, these approaches execute LLM and SSM sequentially, which means the LLM has to wait for several steps of SSM inference before conducting a verification, and vice versa the SSM has to wait for the LLM verification before starting the next round of decoding. The tightly coupled execution of SSM and LLM results in significant idle time during the decoding and verification, hindering further improvement of LLM inference.

%（备注） 1.大模型和小模型参数量差距巨大，存在能力代沟
% (备注）2.小模型的推理步数影响大模型一次验证推出的token数，然而不同模型 不同推理数据集 不同超参设置 均会影响最佳的推理步数 导致最优推理步数搜索空间巨大
% (备注）3.现有的部署方法需要分出GPU对小模型进行单独部署，导致GPU计算资源利用低效；且大小模型的执行过程仍是顺序的，会造成等待 落到计算资源利用低效 怎么组织？

% 可以看出，想要进一步提升speculative decoding方法加速大模型推理需要解决以下几个challenge。首先，LLM和SSM参数量差距巨大（一般在100-1000x），存在能力代沟。如何使SSM更好的近似LLM推理结果是speculative decoding提高SSM acceptance rate迫切需要解决的问题。此外，SSM的speculation length也显著影响着speculative decoding的性能。speculation length越长，LLM一次验证通过的长度也就越长，但同时LLM执行时的计算量也在增加，导致验证的时间增加。最佳的speculation length设置又与不同模型，不同数据集，不同超参数设置（如batch size）密切相关，如何高效搜索上述巨大的优化空间也是提升speculative decoding吞吐的另一个挑战。最后，目前方法将SSM、LLM的执行过程进行了紧耦合且分开部署在不同GPU上，导致大小模型只能顺序执行无法充分利用GPU闲置计算资源。如何实现大小模型的执行解耦，以及大小模型的共置部署是提升speculative decoding推理batch size进而提升推理吞吐的重要挑战之一。
Based on our in-depth investigation of speculative decoding methods, to further accelerate LLM inference, there are several challenges that need to be addressed. Firstly, there is a substantial knowledge gap between SSM and LLM due to the large discrepancy of parameter scale (ranging from 100$\times$$\sim$1000$\times$), which leads to a low acceptance rate from SSM and thus constrains the speedup of LLM inference. \textbf{Challenge 1: How to enable SSM to better align with the inference results of LLM without offsetting its benefit of fast inference?} Additionally, the speculation length of SSM also significantly affects the performance of speculative decoding. A longer speculation length can result in a larger verification cost in each iteration and thus increase the LLM inference time. However, the optimal setting of speculation length varies depending on different models, datasets, and hyper-parameters (e.g., batch size). \textbf{Challenge 2: How to efficiently exploit the vast search space to determine the optimal setting of speculation length?} Finally, the existing methods tightly couple the execution of SSM and LLM, which introduces significant idle time during the interaction of SSM decoding and LLM verification. For example, the sequential execution of SSM decoding and LLM verification forces either SSM or LLM to stay idle when the other is working in progress. \textbf{Challenge 3: How to effectively decouple the execution of SSM and LLM in order to reduce the idle time during decoding and verification?}

% 为了解决这个问题，进一步提升大模型推理吞吐，我们提出了\textit{Minions}, an LLM serving system that accelerates generative LLM inference with efficient pipelined speculative generation. 为了在不引入额外计算开销的情况下提高LLM推理吞吐，\textit{Minions}提出了majority-oriented mechanism，通过使用多个SSMs共同对LLM输出做出预测从而更好的近似LLM的输出。此外，为了更好的在LLM验证开销和其验证通过的token数之间进行权衡，\textit{Minions}提出了adaptive mechanism来动态调整SSM speculation length，从而在不同的serving配置（model、dataset、hyperpara）下获得最优推理性能。在此基础上，我们对LLM和SSMs的执行过程进行了高效解耦并使用模型共置技术，可以在推理过程中支持更大的batch size并高效利用GPU计算资源，进一步提升了LLM的推理吞吐。\textit{Minions}基于vLLM框架进行实现，可以被应用于任意模型框架。We evaluate \textit{Minions}with various general LLM and datasets and compare it with cutting-edge LLM serving systems to demonstrate its effectiveness.
To address the above challenges of speculative decoding with improved inference performance, we propose \textit{Minions}, an LLM inference system that accelerates generative LLM inference with a collective and adaptive speculative generation. To speed up LLM inference without introducing additional computation cost, \textit{Minions} adopts a majority-voted mechanism that leverages the collective wisdom of multiple SSMs to jointly speculate the outputs of LLM. The mechanism can effectively improve the acceptance rate of speculative tokens during LLM verification and thus the inference performance without introducing an overwhelming computation burden. Additionally, to better balance the number of tokens speculated from SSM and the verification overhead of LLM, \textit{Minions} adopts an adaptive mechanism to dynamically adjust the SSM speculation length. The mechanism determines the optimal speculation length to ensure better inference performance across different configurations (e.g., model, dataset, hyper-parameters). Moreover, \textit{Minions} adopts a pipelined execution mechanism that decouples the SSM decoding and LLM verification to effectively reduce the idle time during the inference. The mechanism enables a larger batch size and better utilization of GPU resources, further boosting the inference performance of LLM. \textit{Minions} is implemented based on the vLLM~\cite{kwon2023efficient} and can support main-stream LLMs. We demonstrate the effectiveness of \textit{Minions} by comparing it with cutting-edge LLM inference systems under representative LLMs and datasets.

Specifically, this paper makes the following contributions:
\begin{itemize}
    \item We propose a majority-voted mechanism that leverages multiple SSMs to improve the acceptance rate of speculative tokens during LLM verification with a low computation cost for the inference.
    \item We propose an adaptive mechanism to determine the optimal speculation length of SSMs by efficiently exploiting the search space composed of different LLM configurations for better inference performance.
    \item We propose a pipelined execution mechanism that decouples the execution of SSM decoding and LLM verification to effectively reduce the idle time during the inference with improved throughput.
    \item Based on the above mechanisms, we implement \textit{Minions}, an LLM inference system to accelerate the speculative generation of LLM inference. The experiment results demonstrate that the \textit{Minions} can achieve shorter latency and higher throughput compared to existing systems.
\end{itemize}

The rest of this paper is organized as follows. Section~\ref{sec:motivation} describes the background and motivation of this paper. Section~\ref{sec:methodology} presents the details of our proposed method. Section~\ref{sec:implementation} describes the implementation details. Section~\ref{sec:evaluation} provides the evaluation results compared to the state-of-the-art approaches. We discuss the related work in Section~\ref{sec:relatedworks} and conclude this paper in Section~\ref{sec:conclusion}.
\section{Background and Motivation}
\label{sec:motivation}
In this section, we start by introducing speculative decoding as the background. We then present important observations based on quantitative experiments to illustrate new opportunities for accelerating speculative decoding. We use publicly released LlaMA-160M and OPT-125M as SSMs~\cite{miao2023specinfer}. The remaining experimental setup can be referred to in Section~\ref{subsec:setup}.
% \subsection{LLM Inference}
%简要介绍LLM推理的特点以及现状，主要是把一些重要的概念给出来，例如自回归、prefill、kvcache等，确保后面方法和实现中的一些概念前面都说过。本小节长度可与下一小节差不多。
%The generated output token is fed back into the model to generate the next token.

\subsection{Speculative Decoding}
The fundamental idea of speculative decoding is to use a small speculative model (SSM) to predict several subsequent tokens in advance and then feed them into its counterpart, the large language model (LLM) to verify its speculations. If the speculations from SSM are consistent with LLM, the LLM accepts these speculative outputs. Since the actual token generations are carried out with the SSM that is an order of magnitude faster than LLM, the more tokens from the SSM that are verified by the LLM, the more it accelerates the LLM inference. 

%The process of speculation decoding is shown in Algorithm~\ref{alg:sd}. 

Specifically, in each iteration, the SSM predicts $s$ subsequent tokens ($x_1, x_2, ..., x_s$), and LLM calculates the logit $o_i(x)$ corresponding to each token in parallel. When the computation from both SSM and LLM is complete, the outputs from SSM are verified as follows. For each token, if the logit of SSM ($q_i(x)$) is less or equal to that of LLM ($o_i(x)$), the token will be accepted. Otherwise, the token will be rejected with probability $1-\frac{o_i(x)}{q_i(x)}$ and sampled again from an adjusted distribution of $norm(max(0, o_i(x)-q_i(x)))$. Once the rejection occurs, all tokens after this token will be discarded. Therefore, the efficacy of speculative decoding largely depends on the ability of SSM to accurately predict the outputs of LLM. Moreover, if every token generated by SSM is accepted, an extra token will be sampled using $o_{i+1}(x)$. The above process is repeated until the LLM finishes generating all tokens.

\subsection{Model Capability Gap Between SSM and LLM}
\label{subsec:motivation1}
%表x展示了两组模型在不同数据集下SSM speculation length为8时LLM的平均acceptance rate。
%TODO：此处是否可以补充不同 speculation length的acceptance rate。预期：随着speculation length增加，acceptance rate在下降，但spec的本质还是希望speculation length越长越好，所以还是可以说明大小模型能力有差距。

\begin{figure}[htbp]
    \centering
    \includegraphics[width=\columnwidth]{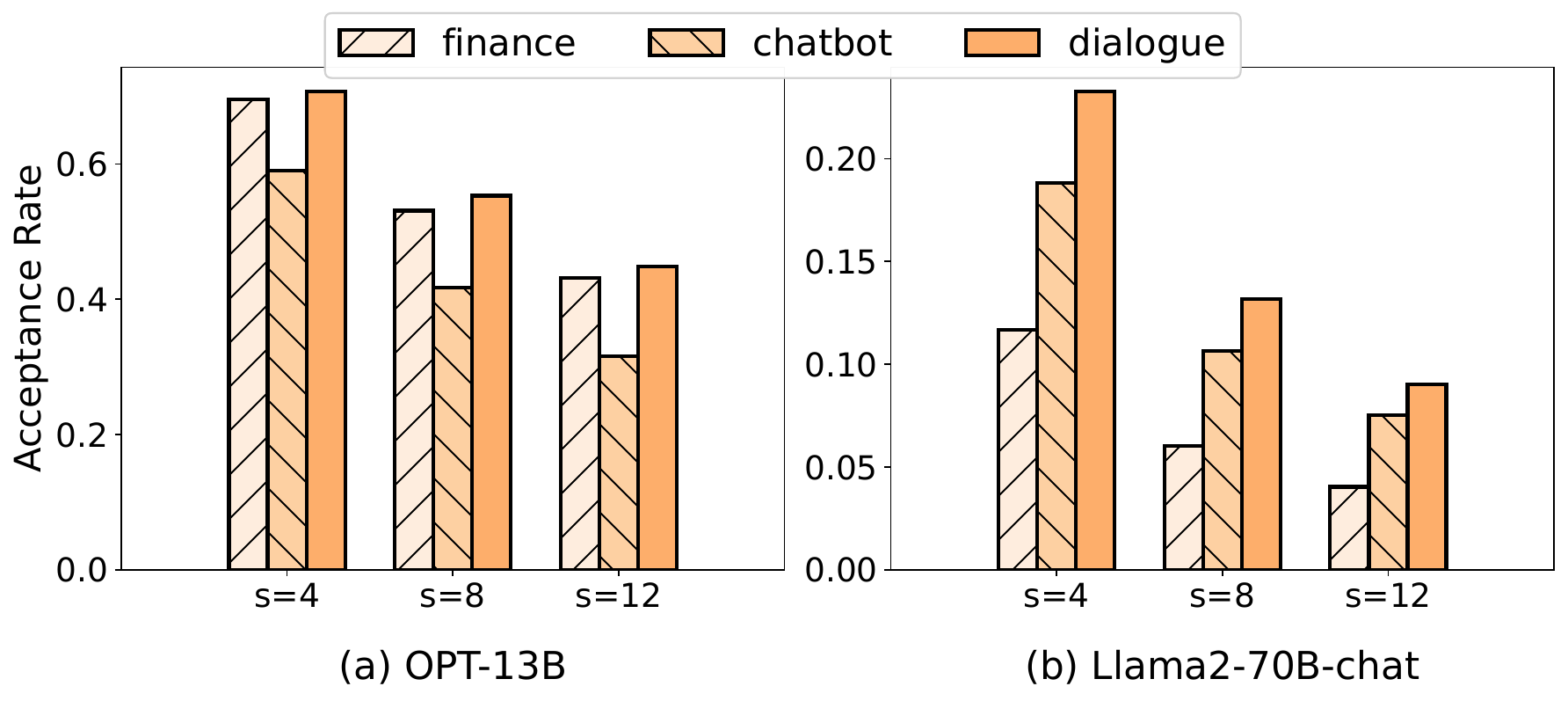}
    \caption{The acceptance rate of models with different speculation lengths and datasets.}
    \label{fig:motivation_3_1}
\end{figure}

%对于OPT-13B-OPT-125M，其在finance和chatbot数据集上表现较优，但在dialogue上略逊一筹。对于Llama2-70B-Llama-160M，平均acceptance rate都很低。SSM acceptance rate较低的原因在于speculative decoding一般使用参数量较小的模型以减少其自回归解码的成本，LLM和SSM在参数量上的巨大差异导致它们的token distribution难以对齐。此外，实验也可以看出在不同数据集上SSM的能力也存在差异，这是由于SSM使用了不同的数据集进行finetune，导致SSM在不同LLM任务上能力具有差异性。可以看出，上述的参数量差异、finetune数据集差异导致了LLM和SSM在预测能力上存在明显能力代沟。如何使SSM更好的近似LLM推理结果是提高SSM acceptance rate迫切需要解决的问题。
%TODO：此处能否补充使用了不同模型架构的SSM实验结果吗？
Figure~\ref{fig:motivation_3_1} presents the average acceptance rates of speculative decoding for two pairs of LLM-SSM models under varying speculation lengths across different datasets. The model pair \textit{OPT-13B-OPT-125M} exhibits a superior acceptance rate on \textit{finance} and \textit{dialogue} datasets, however performs slightly poor in the case of \textit{chatbot} dataset. Whereas for the model pair \textit{Llama2-70B-chat-Llama-160M}, the average acceptance rate is consistently low, especially on the \textit{finance} dataset. The lower acceptance rate can be attributed to the large discrepancy in parameter scale between LLM and SSM, which leads to significant challenges in aligning their token distributions. Moreover, Figure~\ref{fig:motivation_3_1} also shows the varying capability of SSM across different datasets. This is because the SSMs are fine-tuned on different datasets, which leads to varying performance in various LLM tasks. In sum, the difference in parameter scale and fine-tuning dataset leads to a notable capability gap between LLM and SSM that constrains the acceptance rate of SSM and thus the inference performance of speculative decoding.
%It can be observed that the differences in parameter sizes and fine-tuning datasets mentioned above have resulted in a notable performance gap between LLM and SSM in terms of predictive capabilities. Therefore, how to achieve better alignment between SSM and LLM in the current landscape of LLM applications is an important issue to address.

\begin{figure*}[htbp]
    \centering
    \includegraphics[width=0.95\linewidth]{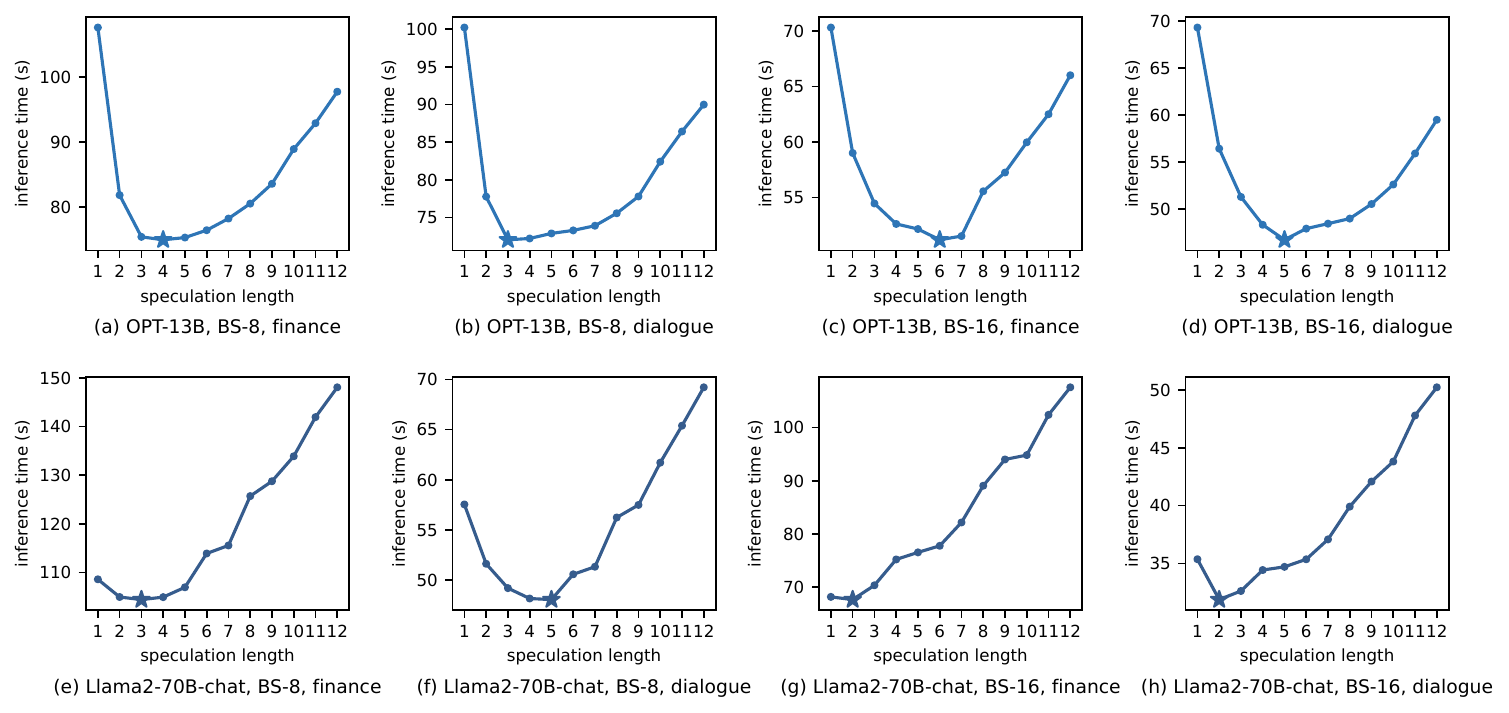}
    \caption{The inference time of the LLM under different speculation lengths, datasets, and batch size.}
    \label{fig:motivation_2}
\end{figure*}
%已有的工作[SPEC]尝试fine-tune出多个SSM，并实现了LLM并行验证，从而联合使用多个SSM以弥合LLM和SSM之间的能力代沟。然而，并行验证多个SSM的输出结果将导致LLM验证计算量激增，offset了SSM推理的性能优势，最终影响LLM的推理性能。因此，如何在不增加LLM验证负担的情况下提高SSM的acceptance rate，是进一步提升LLM推理性能的重要挑战之一。
To fill the capability gap, existing work~\cite{miao2023specinfer} adopts several SSMs to generate multiple speculative outputs simultaneously, which are then verified by the LLM in parallel. However, concurrent verification of multiple SSM outputs can lead to a sharp increase in the computation overhead for LLM, which can quickly offset the performance advantage of SSM inference and ultimately degrade the inference performance of LLM. \textbf{Observation 1: To further improve the performance of speculative decoding, it is challenging to take advantage of the increased capability of multiple SSMs without burdening the LLM verification.}

\subsection{Vast Search Space for Determining Optimal Speculation Length}
\label{subsec:motivation2}

%图x展现了在不同设置（model、dataset、hyperpara）下LLM推理完成时间随speculation length变化的曲线。在这些图中，星号标注的点表示不同配置下的最优speculation point。
%基于对结果的分析，我们得到了两个key observation。首先，无论数据集、模型以及batch size设置如何，随着SSM的speculation length增加，LLM的推理时间都呈现先减少再增加的趋势，且存在最优的speculation length。这是由于随着speculation length的增加，LLM接收的token长度也同样增加，减少了LLM verification的次数，从而提高了speculation decoding的效率。然而与此同时，LLM verfication的计算量也在增加，导致验证一次的时间增长。二者之间存在tradeoff，在不导致验证时间激增的情况下的最长speculation length即为最优。
Figure~\ref{fig:motivation_2} presents the LLM inference time under different speculation lengths and model configurations (datasets, hyper-parameters). The asterisk in each figure annotated the optimal setting of speculation length for each model configuration. As shown in Figure~\ref{fig:motivation_2}, regardless of the dataset and hyper-parameters (e.g., batch size), the inference time of LLM exhibits a decreasing-then-increasing trend when increasing the speculation length of SSM. An optimal setting of speculation length exists that minimizes the LLM inference time. The reason is that a larger speculation length can reduce the number of LLM verifications, which effectively improves the performance of speculation decoding. However, when the speculation length reaches a turning point (asterisk in Figure~\ref{fig:motivation_2}), the computation cost of LLM verification rises significantly, which offsets the benefit of reduced LLM verifications. %There exists a tradeoff between these factors, and the optimal speculation length is the one that maximizes efficiency without causing a sharp increase in verification time.

%其次，我们还发现最优的(the best-performing) speculation length 依赖于模型、数据集的选择以及batch size的设置。For example, with a batch size of 16 on OPT-13B execution, the optimal speculation length is 8 for finance and 5 for dialogue. Using a batch size of 8 with finance, the optimal speculation length is 7 for OPT-13B and 3 for the Llama2-70B-chat model. Similarly, using the dialogue on Llama2-70B-chat execution, the optimal speculation length is 5 with a batch size of 8 and 2 with a batch size of 16. Therefore, the optimal choice for the speculation length involves a vast search space. Navigating such a large search space ends up with a prohibitively long execution time with significant costs, which is unaffordable. Therefore, it is challenging to design an effective speculation length decision mechanism for efficient LLM serving without incurring additional overhead.
Moreover, we notice there are several factors such as dataset and batch size that determine the optimal speculation length. For example, with a batch size of 16, the optimal speculation length of OPT-13B is 6 and 5 under \textit{finance} and \textit{dialogue} datasets, respectively. Similarly, under \textit{dialogue} dataset, the optimal speculation length of Llama2-70B is 5 and 2 with batch size of 8 and 16, respectively. The above factors actually form a vast search space that is prohibitive to exploit exhaustively in order to determine the optimal speculation length. \textbf{Observation 2: An optimal speculation length exists for speculative decoding, however, it is challenging to exploit the search space efficiently to identify the optimal setting.}

%Therefore, the optimal choice for the speculation length involves a vast search space. Navigating such a large search space ends up with a prohibitively long execution time with significant costs, which is unaffordable. 

%注意：这块的逻辑是第一段先说明存在最优点，第二段说明找到最优点搜索空间很大，难以离线确定，需要运行时机制。最后要总结一句话与intro呼应：如何在speculative decoding场景下高效搜索上述巨大的优化空间也是提升LLM推理性能的另一个重要挑战。

\subsection{Tightly-Coupled Speculative Execution Hinders Inference Performance}
\label{subsec:motivation3}

%注意：关于serving和inference的用词，建议尽量使用inference，因为我们最后也没有在线serving的验证场景，还是使用inference更稳妥。
% 先说紧耦合需要pipeline执行，再说共置，因为优化推理性能更重要，降低成本是额外收益。
%\textbf{Idle Waiting for Tightly-Coupled SSM-LLM Execution - }
% 如图x上半部分所示，现有的speculative decoding方法LLM和SSM会顺序执行。即SSM获取一个batch的数据推理speculation length步后交给LLM进行验证。LLM验证后，SSM再继续下一个batch的推理，以此往复。这导致LLM和SSM执行过程中会有大量的空闲等待时间（considerable idle waitings）, 没有充分利用GPU资源的同时也导致执行非常低效。具体来说，如图x下半部分所示，如果能够将SSM和LLM的执行进行高效解耦，SSM在执行完一轮推理后即可开始下一轮推理。同理，LLM验证完一次后也可以直接开始下一轮的验证。在最优情况下，SSM和LLM可以并发执行从而隐藏掉SSM执行开销，显著降低系统的推理延迟。此外，在解耦后，通过将大小模型进行共置执行，可以充分利用空闲的GPU计算资源，从而进一步节省部署成本。因此，如何在不影响大模型执行效率的情况下实现大小模型的高效执行解耦是提升LLM推理性能的重要挑战之一。

\begin{figure}[htbp]
    \centering
    \includegraphics[width=0.9\columnwidth]{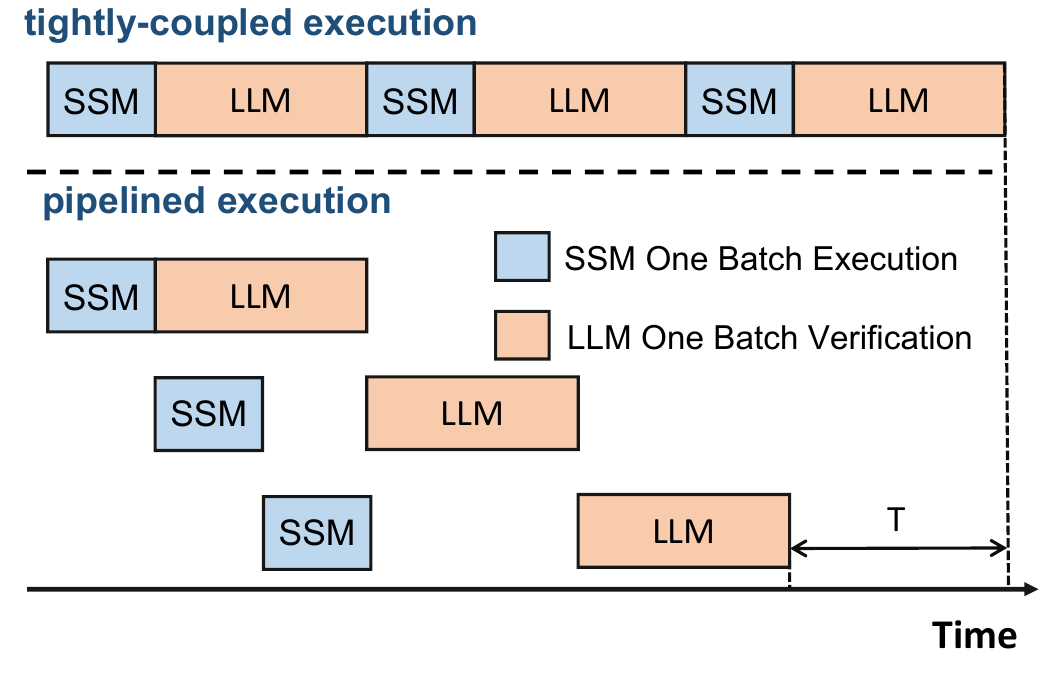}
    \caption{Performance potential revealed by pipelined execution. $T$ represents the reduced inference time.}
    \label{fig:motivation_3_2}
\end{figure}

As shown in the upper part of Figure~\ref{fig:motivation_3_2}, the existing speculative decoding methods execute SSM and LLM sequentially. Specifically, SSM receives a batch of input tokens, performs speculative inference for several steps, and sends the speculative outputs to LLM for verification. After the LLM verification, SSM proceeds with speculative inference for the next batch, and the above process repeats until all tokens have been generated. Consequently, due to the sequential execution of SSM and LLM, there is a large portion of idle time for both SSM decoding and LLM verification, constraining the inference performance. However, as shown in the lower part of Figure~\ref{fig:motivation_3_2}, if the SSM and LLM execution can be decoupled efficiently, after completing one batch of speculation, the SSM can immediately proceed to the next batch. Similarly for LLM, after completing one batch of verification, it can proceed to the next batch. Ideally, the pipelined execution of SSM and LLM can overlap the execution cost of SSM, which can effectively reduce the inference time of LLM. Moreover, decoupling the SSM decoding and LLM verification enables more opportunities to consolidate the execution of SSM and LLM on underutilized computation resources of GPU, which helps to reduce the deployment cost of speculative decoding. \textbf{Observation 3: The tightly-coupled design of speculative decoding forces SSM and LLM to execute sequentially, whereas the decoupled design of SSM decoding and LLM verification reveals promising opportunities for improving inference performance and GPU resource utilization.}

%注意：实验结果中一定要把pipeline执行的性能空间体现出来。

%\textbf{Low GPU Utilization for LLM Execution - } 
% 图x展示了在GPU上使用不同模型和数据集进行LLM inference的achieved occupancy。对于Llama2-70B和OPT-13B，它们的平均achieved occupancy分别为12.19%和33.80%。这表明GPU上仍有很大一部分的计算资源没有被充分利用。除了优化单个LLM推理任务本身以提高GPU利用率，同时在单个GPU上部署多个推理任务也是提高GPU利用率的有效方法之一。然而，共置任务会与LLM推理任务进行资源抢占，即干扰LLM推理执行。因此共置SSM和LLM推理任务的选择需要考虑SSM和LLM任务的计算特征，能够在提升资源利用率降低部署成本的前提下，降低对LLM推理性能的影响。
%TODO：有性能干扰的实验吗？
%TODO：

%最后要总结一句话与intro呼应：如何实现大小模型的执行解耦，以及大小模型的共置部署是提升LLM推理性能、降低的speculative decoding部署成本的重要挑战之一。

% \begin{figure}[t]
%     \centering
%     \includegraphics[width=0.9\columnwidth]{fig/motivation_3_1.pdf}
%     \caption{Achieved occupancy of LLM serving with different datasets on GPUs.}
%     \label{fig:motivation_3_2}
% \end{figure}
%

\section{Methodology}
\label{sec:methodology}
\subsection{Design Overview}
\begin{figure*}[t]
    \centering
    \includegraphics[width=0.85\linewidth]{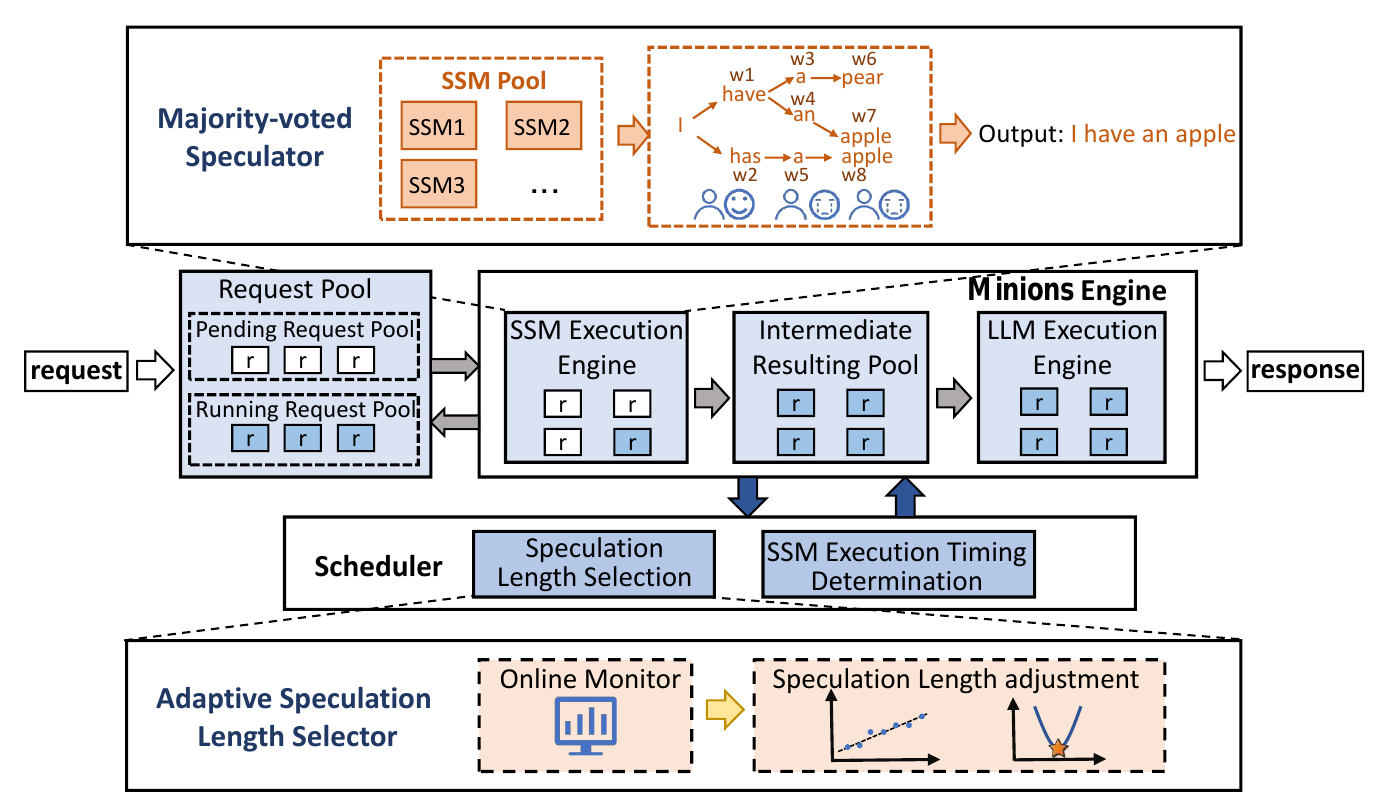}
    \caption{Design overview of \textit{Minions}}
    \label{fig:design_overview}
\end{figure*}

% Minions将新到来的请求放入pending request pool，其负责管理还没有被LLM处理过的请求。此外，对于已经被LLM验证过的请求，我们会将它放入running request pool中，SSM会优先从running request pool中获取请求以便先arrive的请求能够被更早推完。pool会被scheduler进行监控，它负责从request pool中 select a set of requests for SSMs并根据LLM运行时信息通过adaptive speculation length selector模块给出SSM执行的speculation length。其同时还会对inter-result pool进行监控，并从中select a set of requests for LLM verification。在从scheduler获取执行的请求以及speculation length后，SSM execution engine负责SSMs的execution(speculation), 其使用majority-voted **speculator**,通过多个SSMs的weighted vote选出多数认同的SSM输出作为给LLM进行验证的请求。这些请求会被送入中间结果池中等待被LLM执行。在从scheduler获取执行的请求后，LLM execution engine会执行LLM验证过程，验证过程中scheduler中的adaptive speculation length selector模块会使用online monitor监控LLM的验证时间以及验证长度并传递到collector中进行记录。之后根据这些信息对SSMs下次执行的speculation length进行动态调整。对于LLM验证完的每条请求，scheduler会对其进行判断，如果其仍未执行完毕，则会被送入running request pool中。
% TODO：Collector感觉跟online monitor重合了，建议删除或者更加详细说明两个模块的作用
% TODO: 这块overview图不太清晰，最好把request结果的流动也画进去（例如pending/running pool中的request首先经过SSM后得到为speculated request, 再经过LLM后是verified request然后再放回到running pool，优先调度），应该清楚地、直观地看到所有模块间关系或者整个调度执行流。
In this section, we propose \textit{Minions}, an LLM inference system with collective and adaptive speculative generation. As shown in Figure~\ref{fig:design_overview}, newly arrived requests are placed into the pending request pool to wait for execution. \textit{Minions} proposes the \textit{Minions Engine} to execute the requests with a pipelined execution mechanism. It utilizes an \textit{intermediate resulting pool} to store outputs of SSM which decouples the SSM decoding and LLM verification. To ensure the efficiency of the pipeline, \textit{Minions} leverages the \textit{scheduler} to track the \textit{intermediate resulting pool} for determining the timing of SSM execution. Simultaneously, the \textit{scheduler} also dynamically determines the speculation length for SSM execution based on LLM runtime information through the \textit{adaptive speculation length selector}. With speculation length and execution timing from the \textit{scheduler}, \textit{Minions Engine} speculates the results of requests with the SSM execution engine. Specifically, it leverages a \textit{majority-voted speculator} to leverage the collective wisdom of multiple SSMs to determine the majority-approved SSM output. These majority-approved outputs are then placed into the \textit{intermediate resulting pool} to wait for the LLM verification. \textit{Minions Engine} leverages the LLM execution engine to conduct the verification. If the request remains incomplete after this verification round, it will be moved to the running request pool to wait for the next execution.
\subsection{Majority-voted Speculator}
% 扩展小模型的能力是spec的关键。为了解决Section2中提到的现有spec无法应对多场景的挑战，我们通过知识蒸馏（待确认）的手段根据大模型在各个场景数据集下的输出fine-tune出了能够应对不同场景的不同小模型。然而，即使如此，由于无法预知用户行为，如果放置在spec系统中的小模型和用户场景不匹配，仍会造成xxx。一种naive的做法是将所有场景中的模型均放置于系统中，并将所有模型对一条请求的输出均交由大模型进行验证。尽管这能够在多个场景下保证LLM的acceptance rate，这会导致大模型验证的计算量骤增，最终可能导致推理效率下降。

% Inspired by the majority vote algorithm, we propose the learning-based speculator to address this issue. 我们方法的核心哲学。涉及两个事情每个decision的粒度以及权重怎么算，建树，选择（节点粒度 分支粒度），Selection of Decision Metric，整体流程
% 我们方法的核心哲学：
To solve the challenge mentioned in Section~\ref{subsec:motivation1}, we propose the majority-voted speculator inspired by the weighted majority algorithm~\cite{littlestone1994weighted}. The key point is that each SSM can be regarded as an expert in several fields. To perform as the best expert in hindsight, we follow the majority of experts weighted by their accuracy in the past. Specifically, \textit{Minions} updates the weights over time based on the correctness of outputs from the experts. To achieve this goal, we need to answer two crucial questions:
%如何将待输出的结果转化为 the majority of SSMs
\begin{itemize}
    % 对于每个请求，由于SSMs能力的差异性，它们不一定存在一模一样的输出。Minions需要定义什么样的结果可以代表多数的选择
    \item \textbf{How to define the weighted majority of SSMs?} For each request, due to variations in the capabilities of SSMs, they may not necessarily produce identical outputs. \textit{Minions} needs to define what results can represent the majority of choices.
    % SSM的输出难以和LLM完全一致，因此它的输出没有标准答案。Minions需要定义一个metric去判断whether the output of selected SSM is "correct" or "wrong".
    \item \textbf{How to judge whether the SSM makes a mistake?} The output of SSM is challenging to match exactly with LLM. Therefore, \textit{Minions} needs to define a metric to determine whether the output of the selected SSM is \textit{correct} or \textit{wrong}.
\end{itemize}
To tackle these issues, \textit{Minions} leverages the following techniques, including \textit{tree-based weighted majority decision} and \textit{speculation quality metric}.

\textbf{Tree-based Weighted Majority Decision.}
\begin{figure}[t]
    \centering
    \includegraphics[width=1.0\columnwidth]{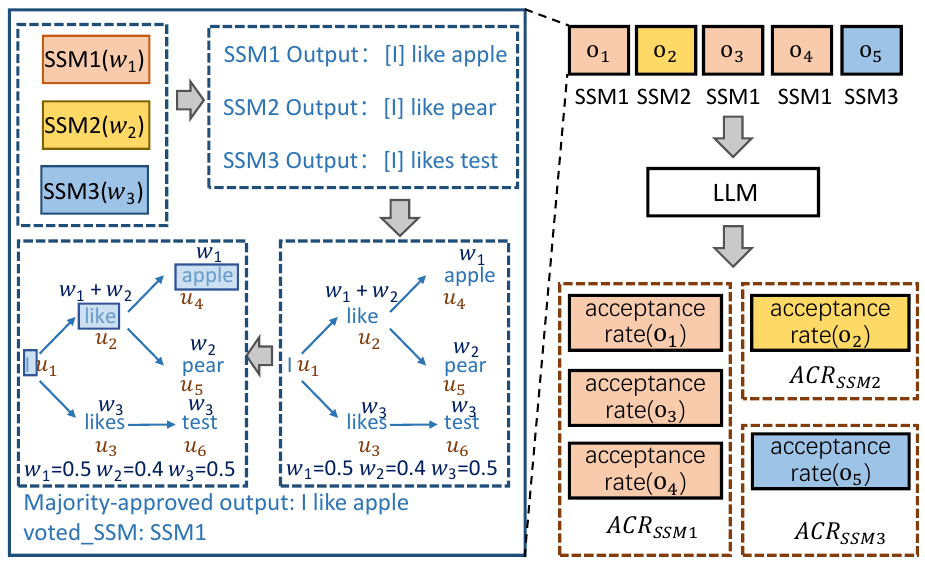}
    \caption{The primary workflow of Majority-voted Speculator.}
    \label{fig:majority}
\end{figure}
% 我们提出使用Tree-cumulative-weight作为评判标准以选择xxx
We leverage a tree structure to represent the output sequences of different SSMs and select the majority-approved output token by token.
% 具体来说，对于每个请求，我们首先将SSM的输出merge为了一棵树。如图x所示，树中的每个节点u代表一个生成的token(tu)，每条以根节点为起点，叶子节点(l)为终点的分支(Bl)代表一个SSM的output。例如，xxx。我们将每个节点的权重定义为输出分支经过该节点的模型权重之和，即xxx。
Specifically, for each request, we first merge the output sequences of SSMs into a tree. As shown in Figure~\ref{fig:majority}, each node $u_i$ in the tree represents a prompt token or generated token (e.g., one word corresponds to one token). The output sequence of each SSM corresponds to a branch ($S_{u_l}$) originating from the root node and terminating at a leaf node ($u_l$). For example, in Figure~\ref{fig:majority}, $S_{u_6}$ represents sequence ($u_1, u_3, u_6$). For each SSM, we assign a weight value ($w_i$). We define the weight of each node ($w_{u_i}$) as the sum of SSMs' weights whose output sequence passes through that node. For example, in Figure~\ref{fig:majority}, $w_{u_2} = w_1 + w_2$ as SSM$_1$ and SSM$_2$ both generate the token associated with 'like'.  
We consider the path that is approved by SSMs with higher weights to be more likely the correct path. As a result, 
after obtaining a weighted tree, the node with the maximum weight is iteratively searched at each tree level starting from the root node to determine the nodes in the majority-approved output, i.e., $v_{i+1} = \mathop{\arg\max}\limits_{u \in c(v_i)}(w_u), S_{final} = S_{v_n}$, where $c(v_i)$ is the set of children of $v_i$ and $v_n$ is a leaf node. For example, in Figure~\ref{fig:majority}, for 'like' and 'likes', as $w_1 + w_2 > w_3$, 'like' is selected. And for 'apple' and 'pear', as $w_1 > w_2$, 'apple' is selected. Thus $S_{u_4}$ is the majority-approved output. It is worth mentioning that the majority-approved output only represents the output of one SSM, and we regard this SSM as $voted\_SSM$. 

%由于speculative sampling的性能强依赖于LLM对SSM推理结果的acceptance rate（i.e. LLM verified length /speculation length）。我们使用average acceptance rate作为判定SSM是否出错的评判指标。具体来说，在通过上述方法得到majority decision后，我们会记录每个decision(r)对应的SSM(d(r) = SSM)。大模型对一个batch的majority decision进行验证后，我们对这次验证中涉及到的SSM求取它的acceptance rate的平均值。即
%$ACR_{SSM} = avg(accept\_rate(r)), d(r) = SSM$.
%通过将ACR_{SSM}与设定的阈值进行比对确定对SSM施加奖励或惩罚。
\textbf{Speculation Quality Metric.} The performance of speculative decoding heavily relies on the acceptance rate of LLM for SSM speculation results (i.e., correctly verified token length/speculation length). As a result, we leverage $SSM\ average\ acceptance\ rate (ACR)$ as the speculation quality metric to assess the performance of $voted\_SSMs$ in the current scenario. Specifically, after validating majority-approved outputs ($o_i$) with LLM, we record the acceptance rate for each output. We denote $o_{SSM_i}$ as the set of $o_i$ with $SSM_i$ as their $voted\_SSM$. For example, in Figure~\ref{fig:majority}, $o_{SSM_1} = \{o_1, o_3, o_4\}$. And $ACR_{SSM_i}$ is defined as the average acceptance rate of all majority-approved outputs in $o_{SSM_i}$, i.e., Equation~\ref{eq:ACR_e}.

\begin{equation}
    ACR_{SSM_i} = \frac{\sum_{o \in o_{SSM_i}}{acceptance\ rate(o)}}{|o_{SSM_i}|}
    \label{eq:ACR_e}
\end{equation}

%LLM验证并对权重进行更新的整体流程如算法x所示。当SSM执行完一个batch后，
In summary, our holistic process is as follows. Once SSMs finish executing a batch of requests, \textit{Minions} will make a tree-based weighted majority decision for each request as discussed above. Meanwhile, \textit{Minions} will also record the $voted\_SSM$ that corresponds to the majority-approved output. Then this output will wait to be fed into the LLM for verification. Once LLM finishes verifying a batch of outputs, $ACR$ of SSMs will be calculated and will be used to update the weights of SSMs. Specifically, we set a \textit{reward\_threshold} and a \textit{punish\_threshold}, once the $ACR$ of the SSM exceeds the \textit{reward\_threshold} or falls below the \textit{punish\_threshold}, we will provide rewards or punishments by multiplying the weight by the corresponding factor.

\subsection{Adaptive Speculation Length Selector}
% 为了解决Section2中最优speculation length选择搜索空间大，开销高的问题，我们首先对speculation length和Minions推理完成时间之间的关系进行了分析。 Table x lists the notations that we use in our modeling. 
To address the challenge of the vast search space for determining the optimal speculation length as discussed in Section~\ref{subsec:motivation2}, we first analyze the relationship between speculation length and the inference time of \textit{Minions}. Table~\ref{table:notation} provides the notations used in our modeling.
\begin{table}[htbp]
    \centering
    \footnotesize
    \caption{Notations used for modeling the inference runtime of \textit{Minions}.}
    \begin{tabular}{|c|c|}
      \hline
      $T_{LLM}$ & Total inference time of the LLM\\
      \hline
      $T_{SSM}$ & Total inference time of the SSM\\
      \hline
      $N$ & Total number of tokens LLM needs to generate\\
      \hline
      $b_{LLM}$ & The batch size of LLM\\
      \hline
      $b_{SSM}$ & The batch size of SSM\\
      \hline
      $s$ & Speculation length\\
      \hline
      $vl(s)$ & \makecell{Average number of SSM generated tokens \\that are verified as correct by LLM}\\
      \hline
      $t_{LLM}(b_{LLM}, s)$ & Per step inference time of LLM\\
      \hline
      $t_{SSM}(b_{SSM})$ & Per step inference time of SSM\\
      \hline
    \end{tabular}
    \label{table:notation}
\end{table}

% Minions的整体执行时间受到两部分影响：(1) SSM的执行时间(Tssm)。(2)LLM的执行时间(Tllm)。
The total inference time of \textit{Minions} is influenced by two parts: (1) the inference time of SSM ($T_{SSM}$), and (2) the inference time of LLM ($T_{LLM}$). 
%As concurrent pipelined execution is employed for SSM and LLM in \textit{Minions} (details will be shown in Section~\ref{subsec:pipeline}), we formulate the total execution time as Equation~\ref{eq:total_execution_time}.
% \begin{equation}
%     T_{total} = \max(T_{SSM}, T_{LLM})
%     \label{eq:total_execution_time}
% \end{equation}
% 我们假设小模型执行一步推理的时间为tssm，执行speculation length(s)步后交给大模型进行一次验证。大模型验证一次的时间和平均验证通过的步数均依赖于s，分别记为tllm(s)和vl(s)。假设大模型需要生成的全部token数记为N，那么总共的验证次数为N/vl(s)。Tssm和Tllm分别为：
The time for the SSM to execute one speculative step is related to SSM batch size ($b_{SSM}$), denoted as $t_{SSM}(b_{SSM})$ and, after performing speculation for a length of $s$ steps, the results are handed over to the LLM for verification. The time required for the LLM to verify once depends on $s$ and LLM batch size $b_{LLM}$, denoted as $t_{LLM}(b_{LLM}, s)$. The average number of tokens that are verified as correct by LLM depends on $s$, denoted as $vl(s)$. Assuming the total number of tokens LLM needs to generate is $N$, the overall verification rounds are given by $\frac{N}{vl(s)}$. Therefore, we can formulate the inference time of the LLM and SSM as Equation~\ref{eq:ssm_time} and Equation~\ref{eq:llm_time}, respectively.

\begin{equation}
    T_{SSM} = \frac{N}{vl(s)} \times t_{SSM}(b_{SSM}) \times s
    \label{eq:ssm_time}
\end{equation}

\begin{equation}
    T_{LLM} = \frac{N}{vl(s)} \times t_{LLM}(b_{LLM}, s) 
    \label{eq:llm_time}
\end{equation}

% 由于SSM和LLM之间的参数量差距，tssm*s小于tllm，使得Tssm小于Tllm。由于Minions采用pipeline执行，更长的Tllm主导着系统的执行时间。因此为了提高Minions的推理效率，需要选择合适的s使得Tllm尽可能小。由于N仅与输入相关，与s无关，因此我们可以将公式进行进一步的简化，即：
Given the large discrepancy of parameter scale between SSM and LLM, $t_{SSM}$ is shorter than $t_{LLM}$, leading to a shorter $T_{SSM}$ compared to $T_{LLM}$. As pipelined execution is employed for SSM and LLM in \textit{Minions} (details will be shown in Section~\ref{subsec:pipeline}), the longer $T_{LLM}$ dominates the total inference time. As a result, to optimize the inference efficiency of \textit{Minions}, the key lies in selecting an appropriate $s$ to minimize $T_{LLM}$. Since $N$ is solely input-dependent and unrelated to $s$, we can further simplify the equation as:

\begin{equation}
    \begin{split}
    s_{optimal} &= \mathop{\arg\min}\limits_{s \in \mathbb{N}}(\frac{N}{vl(s)} \times t_{LLM}(b_{LLM}, s))\\
    &= \mathop{\arg\min}\limits_{s \in \mathbb{N}}(\frac{t_{LLM}(b_{LLM}, s)}{vl(s)})
    \end{split}
    \label{eq:llm_time_new}
\end{equation}

% 基于建模，我们出了一套启发式搜索算法，通过收集模型运行时信息动态调整speculation length。As shown in algorithm 1, 在LLM每次验证时，SpcOpt会使用online monitor对验证一次的时间(t_llm)以及每个请求验证通过的token数进行记录，并计算出验证的平均值(vl)。在得到t_llm和vl后，monitor将t_llm/vl传递给collector进行保存(line b1 ~ b2)。在将s从s1调整为s2后，如图x所示，中间结果池中还存在没来的及被LLM处理的s为s1的请求，因此我们无法立刻通过LLM的运行时信息对这次调整的结果进行判断 能够使monitor收集到充分表征s从s1变化到s2过程中的LLM的运行时信息（即从1到3），对selector做出合理的决定十分重要。为了保证这一点，我们设置了speculation length变化频率的阈值decision\_threshold。只有当collector中收集到了大于等于该阈值的信息后才开始进行speculation length的调整流程(line b3)。具体来说，我们会对collector中的数据进行线性拟合。（line b4）如果得到的一阶系数大于0，则表示s在经历上次变化后t/l呈现上升趋势。根据我们motivation中的观察，这说明目前s的选择大于最优的s，上次变化的操作过于激进。因此我们会对其施加惩罚，减去惩罚步数。如果一阶系数小于0，则说明t/l呈现下降趋势。同样，根据motivation中的观察，说明目前的s仍有继续增大的空间，因此我们会继续对其增加sreward步（line b5-b6）。通过不断的根据运行时反馈对s进行调整，我们最终的性能能够接近于最优s选择下的性能。（详细说明在Sectionx）

\begin{figure}[htbp]
    \centering
    \includegraphics[width=1.0\columnwidth]{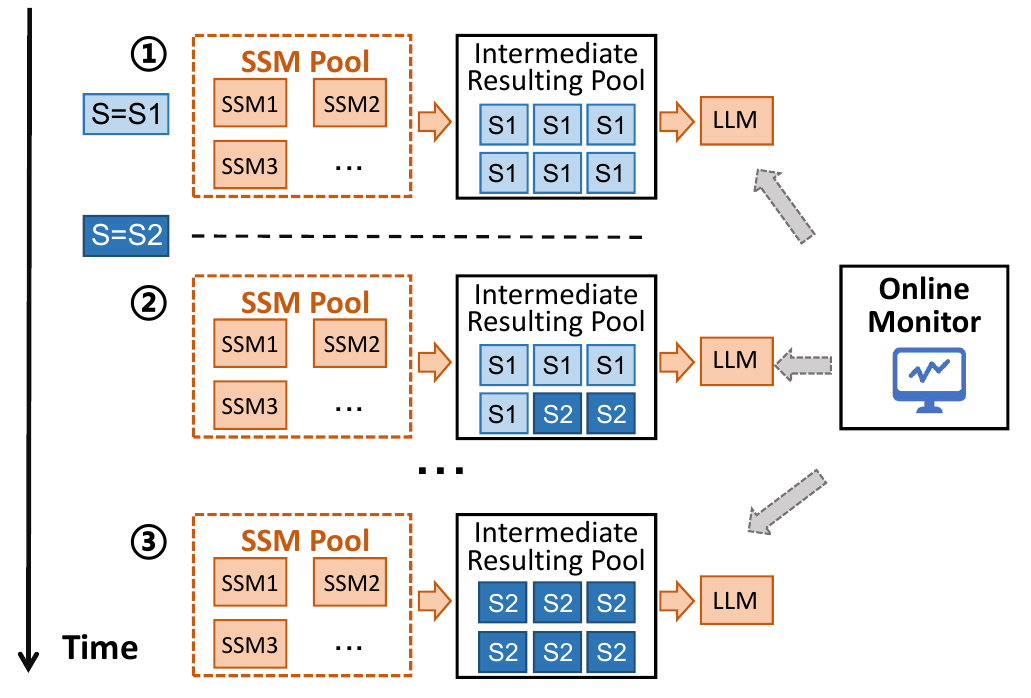}
    \caption{The evolution of requests within the system as $s$ transitions from $s_1$ to $s_2$.}
    \label{fig:monitor}
\end{figure}

Based on this modeling analysis, we propose a heuristic search algorithm that dynamically adjusts the speculation length by monitoring runtime information on LLM verification. Specifically, during each verification of LLM, \textit{Minions} employs an online monitor to record the LLM execution time ($t_{LLM}$) and the number of tokens that are verified as correct for each request. The monitor then calculates the average verified length ($vl$) and records $\frac{t_{LLM}}{vl}$. After adjusting $s$ from $s_1$ to $s_2$, as illustrated in Figure~\ref{fig:monitor}, there are still requests with $s$ set to $s_1$ in the \textit{intermediate resulting pool} that have not been processed by LLM. Consequently, we cannot immediately assess the outcome of this adjustment through the runtime information from LLM. 

To make informed decisions, it is crucial for the online monitor to acquire comprehensive runtime information of LLM, representing the transition of $s$ from $s_1$ to $s_2$ (i.e., from \ding{172} to \ding{174} in Figure~\ref{fig:monitor}). To achieve this, we set a \textit{decision\_threshold} to restrict the frequency of speculation length adjustments. The adjustment only commences when the online monitor has collected information greater than or equal to \textit{decision\_threshold}. Whenever an adjustment process takes place, we perform linear fitting on the monitored data. If the obtained slope is greater than 0, it indicates that, since the last change of $s$, LLM execution has exhibited an upward trend in $\frac{t_{LLM}}{vl}$. Based on our observations from Section~\ref{subsec:motivation2}, this indicates that the current choice of $s$ is greater than the optimal $s$, suggesting that our last change was too aggressive. Therefore, we apply punishment on $s$ by subtracting it from $s_{punish}$. If the slope is less than 0 which indicates a downward trend in $\frac{t_{LLM}}{vl}$. Similarly, our observation from Section~\ref{subsec:motivation2}, suggests that there is still room for further increase in $s$ to further decrease the total inference time. Consequently, we continue to increase $s$ by $s_{reward}$. Through adjustments of $s$ based on runtime feedback, we approach the performance using the optimal $s$ (details in Section~\ref{subsec:step}).

% \begin{algorithm}[htbp]
%     \footnotesize
%     \caption{Adaptive Speculation Length Adjustment}
%     \label{alg:ad}
%     \KwIn{{\color{red}$decision\_threshold$, $s_{penalty}$, $s_{reward}$}}
%     \BlankLine
%     $s \leftarrow init\_speculation\_length$
    
%     \While{$True$}{

%         //\textit{LLM finishes verifying a batch of requests.}
        
%         \If{$LLM finished$}{ \label{line:b1}
%             $t_{LLM} \leftarrow get\_LLM\_veification\_time$
            
%             $vl \leftarrow get\_average\_verified\_length$
            
%             $Monitor\_list.add(\frac{t_{LLM}}{vl})$
%         }\label{line:b2}

%         \If{$Monitor\_list.size > decision\_threshold $}{ \label{line:b3}
%             $coeff \leftarrow Monitor\_list.linearfit$ \label{line:b4}
            
%             \If{$coeff > 0$}{ \label{line:b5}
%                 $s \leftarrow s - s_{penalty}$
%             }
%             \Else{
%                 $s \leftarrow s + s_{reward}$
%             } \label{line:b6}
%             $Monitor\_list.empty$
%         }
%     }

% \end{algorithm}

\subsection{Speculative Generation Pipeline}
\label{subsec:pipeline}
% 如motivation3所示，即使使用了speculation sampling，LLM执行时的GPU仍可能存在低效，这给了我们机会去进一步提升speculation的性能。为了充分的利用宝贵的GPU资源，实现低成本且高效的LLM推理，Minions提出了高效的异步流水线机制，使SSM和LLM并发执行，从而隐藏掉小模型执行引入的额外运行时开销。
As depicted in Section~\ref{subsec:motivation3}, there is a large portion of idle time in tightly-coupled speculative decoding, opening up new opportunities for further improvement in speculation performance. To achieve cost-effective and efficient LLM inference, \textit{Minions} introduces an efficient pipeline mechanism. It concurrently executes SSMs and the LLM and effectively overlaps the additional runtime overhead introduced by the speculation of SSMs.

% SSM和LLM能够并发执行的关键是将他们的执行过程进行解耦。为了解决这个问题，我们提出了中间结果池。SSM执行完一个batch后将speculation result存储在中间结果池中，LLM验证时再从中间结果池获取请求。因此，SSM的执行不会受限于LLM，同时只要中间结果池中仍存在待验证的请求，LLM就可以一直执行。 
The key to enabling pipelined execution of SSMs and the LLM lies in decoupling their execution processes. To solve this issue, we introduce an \textit{intermediate resulting pool}. After inferencing a batch, SSMs store speculation results in the \textit{intermediate resulting pool}, and the LLM accesses the pool for verification. Hence, the execution of SSMs is not constrained by LLM. And as long as there are available requests in the \textit{intermediate resulting pool}, LLM can sustain continuous execution.

% 为了使得流水线可以高效执行，一个需要注意的问题是SSM speculation的速率。避免在LLM执行过程中产生bubble是上述流水线机制实现高吞吐的关键因素。幸运的是，由于SSM和LLM参数量的差距，SSM的执行速度快于LLM，使得LLM每次执行时中间结果池总会存在待处理的请求，因此不会有bubble产生。
%但这同时也会引入另一个问题。如图x所示，SSM的高推理速率会导致LLM无法及时对其推理出的请求进行验证。因此，随着时间的推移，中间结果池会积累大量的请求。而一旦一条请求开始被推理，就会生成KVCache并驻留在显存中。随着LLM推理过程的进行，显存中驻留了大量请求的KVCache，但是只有少部分请求正在被LLM执行。一方面，这会导致系统频繁触发swapping[x]或recomputation[x]以保证显存安全。然而无论哪种显存优化方式，都会产生大量额外的通信或计算开销，导致LLM的推理效率严重受损。另一方面，受限于这种低效的显存管理，LLM推理的batch size也会受限，进而影响LLM的吞吐

\begin{figure}[htbp]
    \centering
    \includegraphics[width=1.0\columnwidth]{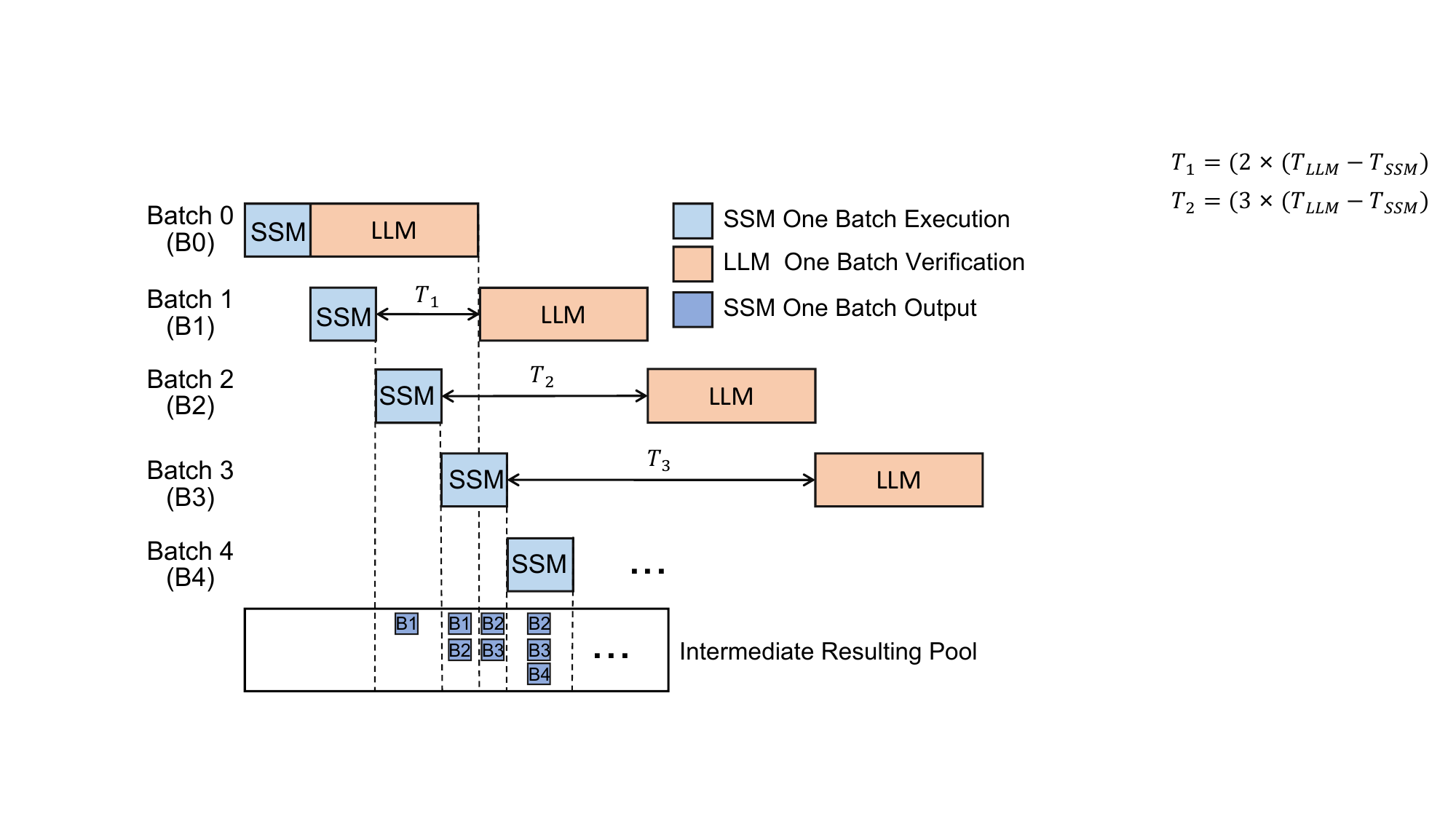}
    \caption{The workflow of the speculative generation pipeline and the status of the \textit{intermediate resulting pool} in the absence of any control. $T_i$ denotes the duration during which requests from Batch $i$ are queued in the \textit{intermediate resulting pool}.}
    \label{fig:pipe}
\end{figure}

To ensure efficient execution of the pipeline, a crucial consideration is the speed of SSM speculation. Avoiding bubbles in the LLM execution process is a key factor in achieving a high throughput pipeline mechanism. Fortunately, due to the large discrepancy in parameter scale between SSM and LLM, the execution speed of SSM is faster than that of LLM. This ensures that there are always pending requests in the \textit{intermediate resulting pool} when LLM is ready to execute, preventing the occurrence of bubbles in the execution flow. Nevertheless, this efficiency comes at the cost of introducing another intricate challenge. As shown in Figure\ref{fig:pipe}, the high inference speed of SSM results in LLM being unable to immediately verify the requests SSM generates. Consequently, over time, the \textit{intermediate resulting pool} accumulates a significant number of pending requests. Once a request is processed, the key and value matrices will be calculated and reside in the GPU memory as KVCache. As the LLM inference progresses, the GPU memory is occupied by a large number of KVCache for requests, but only a small portion of these requests are actively being executed by LLM. On one hand, this leads to frequent triggering of swapping~\cite{huang2020swapadvisor,peng2020capuchin} or recomputation~\cite{kwon2023efficient} to ensure memory safety, which results in severe degradation of the inference efficiency of LLM. On the other hand, constrained by this inefficient memory management, the batch size for LLM inference is also limited, thereby impacting the throughput of LLM.

%为了缓解这个问题，我们通过对中间结果池大小的实时监控对SSM 的speculation speed进行了限制。首先，对于LLM来说，只要在其执行每次推理前中间结果池能够包含bs个请求，pipeline过程中就不会产生bubble。在此限制下，需要尽可能的减少积压在中间结果池的请求数。为了实现这个目标，我们会在SSM推理完一次后获取中间结果池的大小，如果该大小>=llm bs，则小模型会暂停执行，直至某次LLM获取请求后中间结果池中的请求数小于大模型bs。由于处理相同数目的请求，ssm速度高于llm，ssm会在llm验证完成前生成足够数量的请求，（即满足限制）。
To alleviate this issue, we track the \textit{intermediate resulting pool} to restrict the speculation speed of SSMs. Firstly, for LLM, as long as the \textit{intermediate resulting pool} can accommodate requests equivalent to its batch size before each inference execution, there will be no bubbles in the pipeline process. Under this constraint, efforts are made to minimize the accumulation of requests in the pool. To achieve this goal, as soon as a round of SSM speculation finishes, we track the volume of the \textit{intermediate resulting pool}. If the volume is greater than or equal to the batch size of LLM, SSMs pause their execution. However, once it is observed that the volume of the \textit{intermediate resulting pool} falls below the batch size of LLM due to LLM execution, the SSM will execute immediately. Due to the higher speed of SSM compared to LLM in processing an equal number of requests, SSM can generate a sufficient quantity of requests before the completion of LLM verification, thereby satisfying the imposed constraint.
\section{Implementation}
\label{sec:implementation}
We have implemented a system prototype of \textit{Minions} in Python, and built it on top of vLLM~\cite{kwon2023efficient}. The ideas behind \textit{Minions} are general to be adapted to other LLM inference systems, and we leave such engineering efforts for future work.

%  此外，为了将LLM验证失败的token对应的KVCache进行删除，我们对vLLM的BlockManager进行了扩展。在LLM进行验证时记录每个token对应的KVCache的虚拟block id和偏移量。验证后，对于验证失败的token，基于之前记录的虚拟地址删除其对应的memory slot。
The vLLM system is built with the assumption that in the decoding stage, there will be one newly generated token that ties to one KVCache block for each sequence. This assumption won't hold anymore for speculative decoding. To address this issue, we extend the paged attention of vLLM to support taking more than one token as input when KVCache has already existed. Furthermore, to remove the KVCache corresponding to tokens that failed verification in the LLM, we extend the BlockManager of vLLM. Specifically, during LLM execution, for the KVCache of each token, we record its logical block ID and offset. After execution, for tokens that failed verification, we delete their corresponding memory slots based on the previously recorded logical addresses. 

% 为了实现SSM和LLM concurrent inference，我们 utilizes NVIDIA MPS~\cite{mps2012nvidia} which is a practical technique used by several spatial GPU sharing works~\cite{sun2022cognn,choi2022serving}。此外，在Worker管理方面，由于VLLM自带的分布式执行框架Ray无法满足我们将数个Worker分配至同一指定节点的要求，我们使用Python自带的Subprocess框架实现了一套Worker管理与通信机制。具体而言，在启动Worker时，主进程会启动多个子进程作为Worker，并对每个进程建立发送与接受的PIPE。建立的PIPE后续被用于实现主进程与子进程的进程间通信，子进程不断检查PIPE中是否有新的命令，若有则执行相应命令并将结果返回。

% Moreover, to achieve concurrent inference for SSM and LLM, we utilize NVIDIA MPS~\cite{mps2012nvidia} which is a practical technique used by several spatial GPU sharing works~\cite{sun2022cognn,choi2022serving}. In terms of worker management, the built-in distributed execution framework Ray of vLLM falls short of meeting the requirement of assigning multiple workers to the same designated node. To address this issue, we implement a worker management and communication mechanism using multiprocessing of Python. Specifically, when starting workers, the main process initiates multiple subprocesses as workers and establishes a pipe for sending and receiving for each process. The created pipes are subsequently used for inter-process communication. The subprocesses continuously check if there are new commands in the pipe, and if so, they execute the commands and return the results.

In addition, to better align the SSMs with LLM, we fine-tune SSMs with distillation technique~\cite{hsieh2023distilling,smith2022language}. Specifically, we prompt LLM with part of the instructions constructed from Empathetic\_Dialogues Datasets~\cite{rashkin-etal-2019-towards}, Chatbot Instruction Prompts Datasets~\cite{chatbot}, and Finance Alpaca Datasets~\cite{finance} respectively. We then filter out wrong answers and use the correct generated instruction data to fine-tune SSMs.

Moreover, to achieve concurrent inference for SSMs and LLM, we utilize NVIDIA MPS~\cite{mps2012nvidia} which is a practical technique used by several spatial GPU sharing works~\cite{sun2022cognn,choi2022serving}. 
% mps实现共置 由于共置 新的通信方式
\section{Evaluation}
\label{sec:evaluation}
\begin{figure*}[t] 
  \centering
    \includegraphics[width=\linewidth]{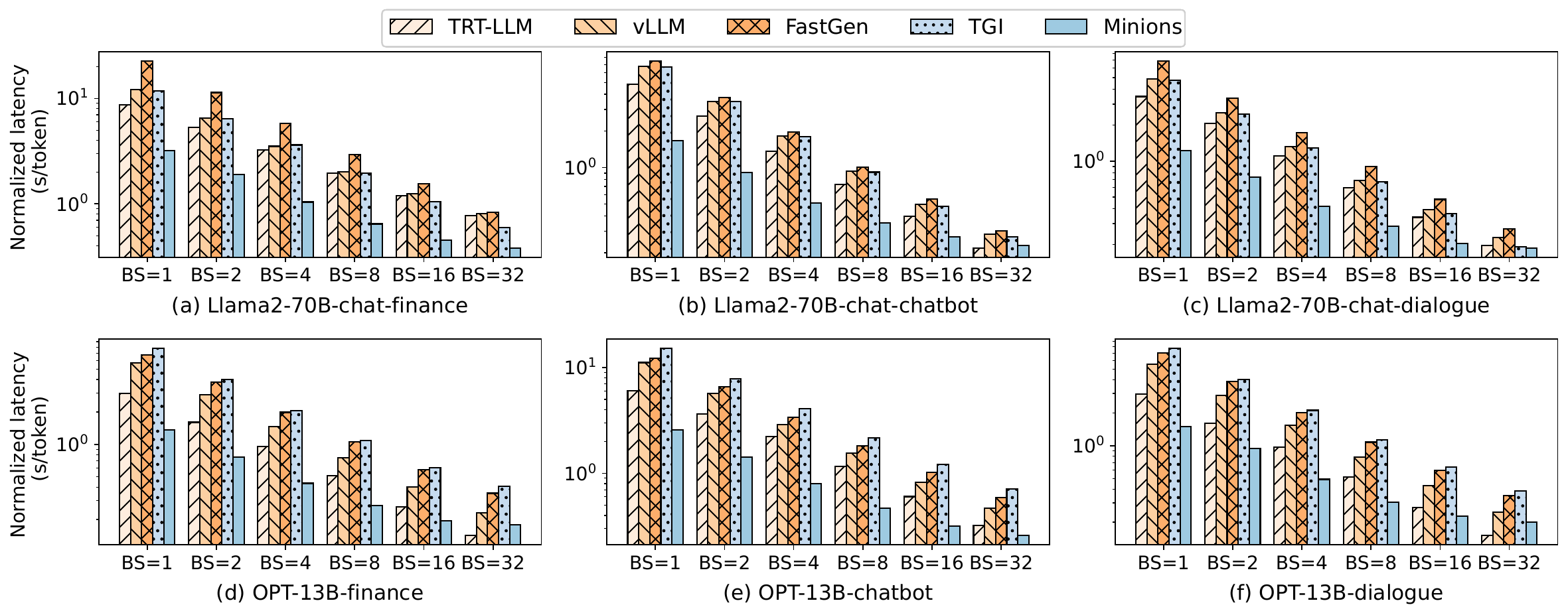}
    \caption{Normalized latency of different LLM inference systems.}
    \label{fig:overall_latency}
\end{figure*}

\subsection{Experiment Setup}
\label{subsec:setup}
\textbf{Hardware and Software Configurations -} 
We run our evaluation on 4 NVIDIA 80-GB A100 GPUs connected over NVLink. The experiments are conducted on Ubuntu 20.04 with CUDA 11.8 and cuDNN v8.9.0. \textit{Minions} is implemented on PyTorch v2.0.1, transformers v4.34.0, xformers v0.0.22 and pydantic 1.10.13. 

\textbf{Models and Datasets -}
Throughout the experiments, we evaluate our system using two representative LLM families: Llama~\cite{touvron2023llama} and OPT~\cite{zhang2022opt}. Specifically, we choose Llama2-70B-chat~\cite{llama2} and OPT-13B~\cite{opt-large} as the LLMs and fine-tune Llama-160M~\cite{llama1} and OPT-125M~\cite{opt-small} to generate SSMs. We evaluate \textit{Minions} on three conversational datasets: Empathetic\_Dialogues Datasets~\cite{rashkin-etal-2019-towards} (\textit{dialogue}), Chatbot Instruction Prompts Datasets~\cite{chatbot} (\textit{chatbot}), and Finance Alpaca Datasets~\cite{finance} (\textit{finance}). To simulate the real-world scenario, for \textit{chatbot} and \textit{finance}, we only use part of the prompt (instruction) field to form our input prompts. And for \textit{dialogue}, we combine partial prompt and utterance field to create the input prompts.

\textbf{Baseline Systems and Comparison Methods -} 
We compare \textit{Minions} with state-of-the-art LLM inference systems including \textit{vLLM}~\cite{kwon2023efficient}, HuggingFace Text Generation Inference (\textit{TGI})~\cite{tgi}, DeepSpeed-FastGen~\cite{fastgen} (\textit{FastGen}) and TensorRT-LLM~\cite{tensorrt-llm} (\textit{TRT-LLM}). All systems serve LLMs with fp16-formatted model parameters and intermediate activations. The difference is that OPT-13B is served with two GPUs utilizing tensor parallelism, while Llama2-70B-chat employs four GPUs with tensor parallelism to ensure memory safety. 

We use prompts from the datasets described above. For each prompt, we let all systems generate a maximum of 128 new tokens. To measure the effectiveness of \textit{Minions}, we use the normalized latency (the mean end-to-end latency of each request divided by its output length) and throughput as the major evaluation metric.

\subsection{End-to-end Performance}
\label{subsec:e2e}
\begin{figure*}[t] 
  \centering
    \includegraphics[width=\linewidth]{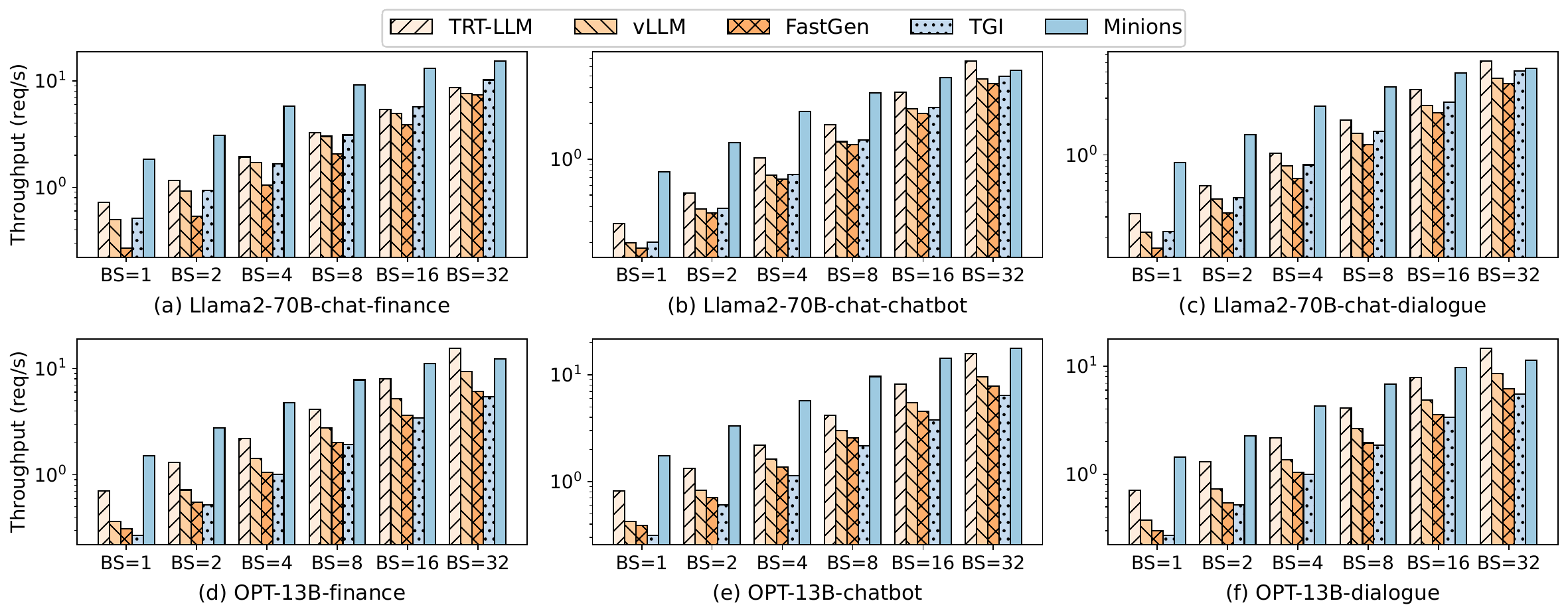}
    \caption{Throughput of different LLM inference systems.}
    \label{fig:overall_throughput}
\end{figure*}

\textbf{Latency -} Figure~\ref{fig:overall_latency} presents the comparison of latency with each LLM inference system. In Figure~\ref{fig:overall_latency}(a)(b)(c), for Llama2-70B-chat with \textit{finance}, \textit{chatbot}, and \textit{dialogue} dataset, \textit{Minions} shows a maximum speedup of 3.11$\times$/2.91$\times$/2.84$\times$, 3.81$\times$/4.06$\times$/3.98$\times$, 7.06$\times$/4.48$\times$/5.62$\times$, and 3.70$\times$/3.99$\times$/3.89$\times$ compared to \textit{TRT-LLM}, \textit{vLLM}, \textit{FastGen}, and \textit{TGI}, respectively. The performance improvement comes from the following aspects. Firstly, \textit{Minions} can leverage the collective wisdom of multiple SSMs to jointly speculate the outputs of LLM without introducing additional computational overhead for LLM. In addition, \textit{Minions} also achieves a good balance between the number of tokens speculated from SSM and the verification cost of LLM, which further boosts the performance. Lastly, with efficient pipelined speculative generation, LLM can continuously perform verification while ensuring that requests are not excessively backlogged in the \textit{intermediate resulting pool}. Note that compared with \textit{vLLM}, in \textit{finance}, \textit{chatbot}, and \textit{dialogue} dataset, \textit{Minions} achieves an average speedup of 3.10$\times$, 2.86$\times$, and 2.70$\times$, which further demonstrates the strength of our work. Figure~\ref{fig:overall_latency}(d)(e)(f) presents similar results for OPT-13B. For \textit{finance}, \textit{chatbot}, and \textit{dialogue} dataset, \textit{Minions} shows a maximum speedup of 2.21$\times$/2.78$\times$/1.99$\times$, 4.22$\times$/4.33$\times$/3.75$\times$, 4.99$\times$/4.74$\times$/4.71$\times$, and 5.70$\times$/5.88$\times$/5.18$\times$ compared to \textit{TRT-LLM}, \textit{vLLM}, \textit{FastGen}, and \textit{TGI}, respectively. As for \textit{TRT-LLM}, our performance is slightly inferior to it when using a batch size of 32 in certain datasets. The reason is that \textit{TRT-LLM} is built on a high-performance inference framework, TensorRT, incorporating highly optimized operators and various additional compilation optimizations. The optimized implementations of models enable it to achieve higher performance than our base system \textit{vLLM}, thereby surpassing ours in certain configurations. However, since the idea behind our work is orthogonal to the system, we believe that there will be additional performance improvement arising by adapting \textit{Minions} to \textit{TRT-LLM}. 

\textbf{Throughput -} Figure~\ref{fig:overall_throughput} presents the comparison of throughput when inferencing two LLM models with different configurations. For Llama2-70B-chat with \textit{finance}, \textit{chatbot}, and \textit{dialogue} dataset, \textit{Minions} shows a maximum speed up of 3.00$\times$/2.71$\times$/2.71$\times$, 3.71$\times$/3.95$\times$/3.88$\times$, 6.84$\times$/4.39$\times$/5.25$\times$, and 3.60$\times$/3.88$\times$/3.81$\times$ compared to \textit{TRT-LLM}, \textit{vLLM}, \textit{FastGen}, and \textit{TGI}, respectively. This is because \textit{Minions} leverages multiple SSMs to collectively assist the inference of LLM and adaptively adjust the speculation length to enhance the verification efficiency of LLM. Compared with \textit{vLLM}, in \textit{finance}, \textit{chatbot}, and \textit{dialogue} dataset, \textit{Minions} achieves an average speedup of 3.02$\times$, 2.76$\times$, and 2.69$\times$, which further demonstrates the performance enhancement achieved by our work. For OPT-13B, similar experimental results are presented. On the \textit{finance}, \textit{chatbot}, and \textit{dialogue} dataset, \textit{Minions} shows a maximum speed up of 2.17$\times$/2.64$\times$/2.03$\times$, 4.18$\times$/4.17$\times$/3.83$\times$, 5.01$\times$/4.66$\times$/4.78$\times$, and 5.65$\times$/5.61$\times$/5.31$\times$ compared to \textit{TRT-LLM}, \textit{vLLM}, \textit{FastGen}, and \textit{TGI}, respectively. For both models, the performance is slightly lower when compared to \textit{TRT-LLM} on certain datasets with a batch size of 32. The reason is also attributed to the high performance of TensorRT, as discussed above.

%\textbf{Acceptance rate -} 
%We evaluate the average acceptance rate of LLMs on various datasets with a batch size of 16. As shown in Table~\ref{table:acceptance_rate}, LLM and SSMs achieve a good alignment. Specifically, for OPT-13B with OPT-125M as SSM, it can achieve an average acceptance rate of 0.76 $\sim$ 0.89, and thus significantly enhance the throughput during LLM execution. For Llama2-70B-chat with Llama-160M as SSM, it can achieve an average acceptance rate of 0.42 $\sim$ 0.54. Compared to OPT, its acceptance rate is slightly lower due to the larger gap in parameter size between SSM and LLM. Nevertheless, compared to other systems that execute sequentially, this acceptance rate can still bring a significant performance improvement to LLM inference.

%\begin{table}[htbp]
%    \centering
%    \footnotesize
%    \caption{The average acceptance rate of \textit{Minions} on various configurations.}
%    \begin{tabular}{|c|c|c|c|}
%      \hline
%       & finance & chatbot & dialogue\\
%      \hline
%      OPT-13B& 0.87 & 0.89 & 0.76\\
%      \hline
%      Llama2-70B-chat & 0.54 & 0.44 & 0.42\\
%      \hline
%    \end{tabular}
%    \label{table:acceptance_rate}
%\end{table}
\begin{figure}[htbp]
  \centering
  \includegraphics[width=0.95\columnwidth]{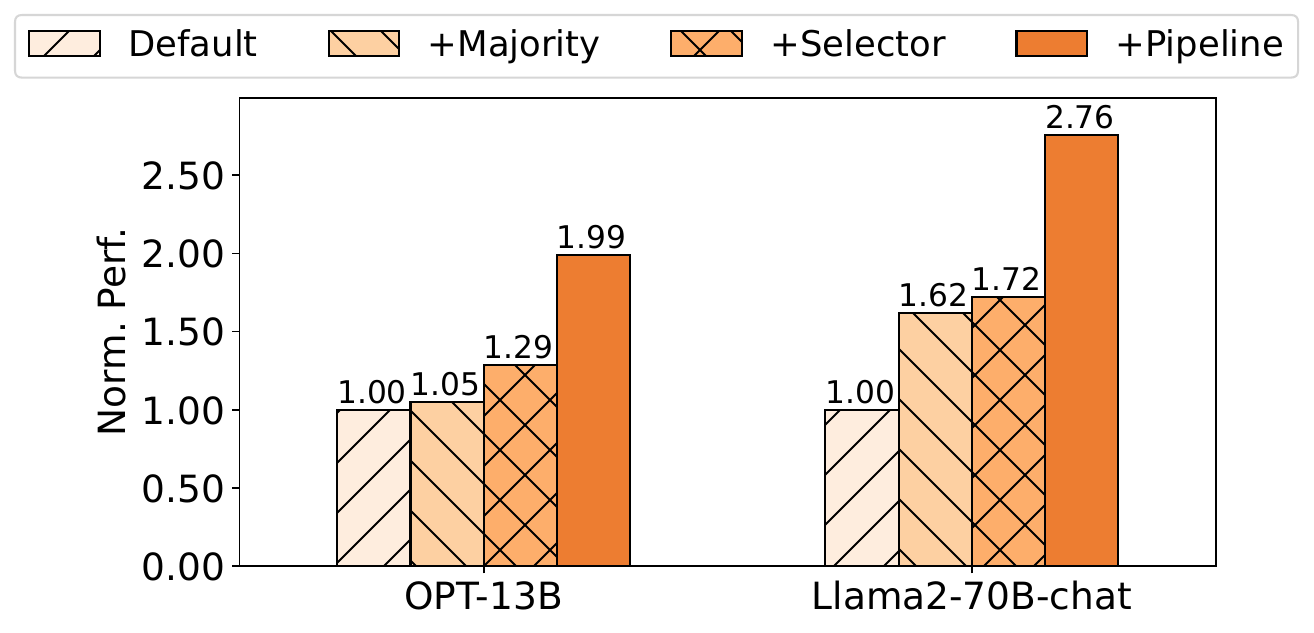}
  \caption{Throughput breakdown of \textit{Minions}. The reported throughputs are normalized by \textit{Default}.}
  \label{fig:ablation}
\end{figure}

\begin{figure*}[htbp]
  \centering
  \includegraphics[width=1.0\linewidth]{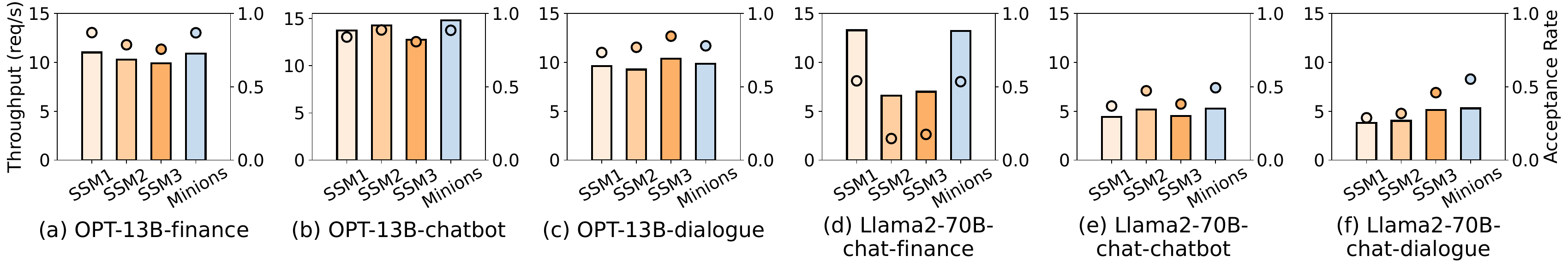}
  \caption{The throughput and acceptance rate of majority-voted speculator compared to the implementations using only one kind of SSM.}
  \label{fig:ablation_1}
\end{figure*}

\begin{figure*}[t]
  \centering
  \includegraphics[width=1.0\linewidth]{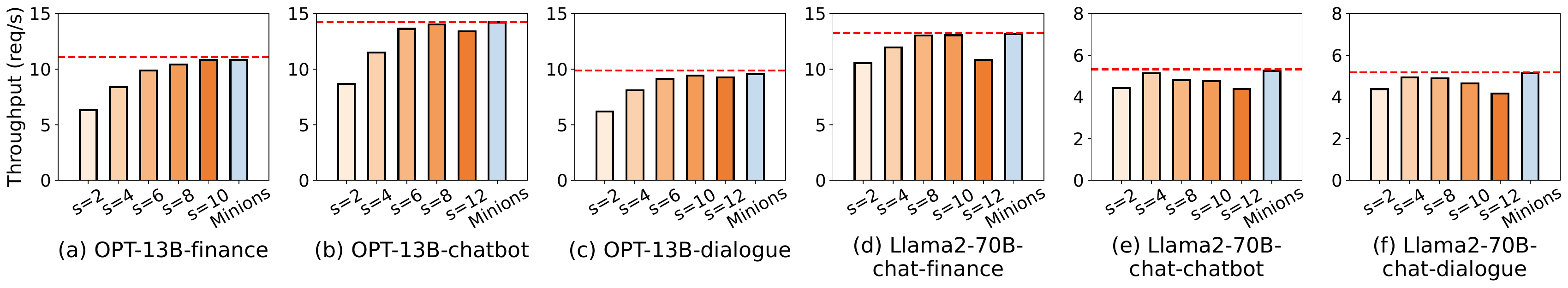}
  \caption{The throughput of adaptive speculation length selector compared to the implementations with fixed speculation length ranging from 2 to 12.}
  \label{fig:ablation_2}
\end{figure*}

\subsection{Ablation Study}
To better understand the contribution of adopted methods in \textit{Minions}, we profile the throughput of \textit{Minions} on \textit{finance} dataset with a batch size of 16 and break down the performance results into three parts. We regard the implementation of traditional speculative decoding mentioned by previous work~\cite{chen2023accelerating} as the baseline ($Default$) and choose the speculation length consistent with it (i.e. 4). We first employ diverse SSMs with a majority-voted speculator (\textit{Majority}) on the basis of \textit{Default}. As shown in Figure~\ref{fig:ablation}, for OPT-13B and Llama-70B-chat, \textit{Majority} can bring an increment of 9.56\% and 59.82\% on throughput. This is mainly because \textit{Majority} unifies the capabilities of multiple SSMs without incurring additional verification overhead. For OPT-13B, the increase is not as substantial as Llama2-70B-chat. The reason is that its employed SSMs exhibit comparable capabilities on the \textit{finance} dataset, thus the combined advantage of their collaboration is weakened.

Based on \textit{Majority}, we further demonstrate the effectiveness of adaptive speculation length selector (\textit{Selector}). Specifically, it delivers extra throughput increments of 35.07\% and 5.91\% on OPT-13B and Llama2-70B-chat, respectively. The main reason is that \textit{Selector} can dynamically adjust the speculation length to bring it closer to the optimum, greatly balancing the verified length and verification cost of LLM. For Llama2-70B-chat, the rise is not significant compared to OPT-13B. The reason is that the default speculation length is very close to the optimum. As a result, only a minimal optimization space remains for \textit{Selector}.

Based on \textit{Majority} and \textit{Selector}, we further consider the effectiveness of the speculative generation pipeline (\textit{Pipeline}). For OPT-13B and Llama-70B-chat, it additionally brings an increase of 34.27\% and 55.37\% on throughput. This is mainly because \textit{Pipeline} decouples the execution of SSM decoding and LLM verification, effectively reducing the idle time during the inference with improved throughput.

% \label{subsec:step}
\subsection{Effectiveness of Design Choices}
\label{subsec:step}
\textbf{Effectiveness of Majority-voted Speculator -}
% we compare Minions with 只使用一种任务上fintune出的SSM。由于xxx，我们将三个SSM中表现最好的一个作为optimal xx， which is demonstrated as the red line.
% 我们将finetune得到的SSM通过比较Minions和只使用一种SSM执行speculation的实现进行对比（）
To demonstrate the effectiveness of our proposed majority-voted speculator, we evaluate \textit{Minions} with the implementations using only one kind of SSM, respectively recorded as SSM1, SSM2, and SSM3. The batch size we use is 16. The detailed evaluation results are shown in Figure~\ref{fig:ablation_1}, where the bars indicate the throughput of each evaluated employment, and the circled dots are the acceptance rates. As shown in Figure~\ref{fig:ablation_1}, in all diverse model and dataset configurations, \textit{Minions} is able to achieve throughput comparable to, or even higher than the optimal throughput in SSM1, SSM2, and SSM3. The reason is that through the majority-voted mechanism, \textit{Minions} can leverage the collective wisdom of multiple SSMs under a dataset by adjusting weights based on runtime feedback. The results of the acceptance rate further demonstrate this point. As shown in Figure~\ref{fig:ablation_1}, for OPT-13B, \textit{Minions} is able to achieve acceptance rates of 0.87, 0.89, and 0.78 on \textit{finance}, \textit{chatbot}, and \textit{dialogue} respectively. For Llama-70B-chat, \textit{Minions} is able to achieve acceptance rates of 0.54, 0.49, and 0.55 on \textit{finance}, \textit{chatbot}, and \textit{dialogue} respectively.

\textbf{Effectiveness of Adaptive Speculation Length Selector -}
To illustrate the effectiveness of our adaptive mechanism to dynamically adjust the speculation length ($s$) of SSMs, we compare \textit{Minions} with the implementations with fixed $s$ ranging from 2 to 12. The batch size we use is 16. Particularly, we leverage the red line to represent the highest throughput across all steps, regarding it as the optimality. As shown in Figure~\ref{fig:ablation_2}, in all diverse model and dataset configurations, \textit{Minions} consistently achieve high throughput that closely approaches the optimality and outperforms others. The result demonstrates that through several rounds of adjustments, \textit{Minions} is capable of correctly selecting the optimal speculation length, and therefore greatly enhances inference throughput. 

% \textbf{Effectiveness of Pipelined Speculative Generation-}
% For the ablation study of the effectiveness of pipelined speculative generation, we evaluate \textit{Minions} with the implementation that sequentially executes SSMs and LLM (Sequential). As shown in Figure~\ref{fig:ablation_3}, \textit{Minions} achieves 1.56 $\sim$ 1.69 $\times$ and 1.58 $\sim$ 1.62 $\times$ speedups on OPT-13B and Llama2-70B-chat, respectively. This is mainly because with the pipelined speculative generation, \textit{Minions} efficiently overlaps the execution of SSMs and LLM and carefully controls the volume of \textit{intermediate resulting pool} and thus improves inference throughput. 
% \begin{figure}[t]
%     \centering
%     \includegraphics[width=\columnwidth]{fig/ablation_3.pdf}
%     \caption{xxx.}
%     \label{fig:ablation_3}
% \end{figure}

\subsection{Comparison with Speculative Decoding}
\begin{figure}[htbp]
  \centering
  \includegraphics[width=0.85\columnwidth]{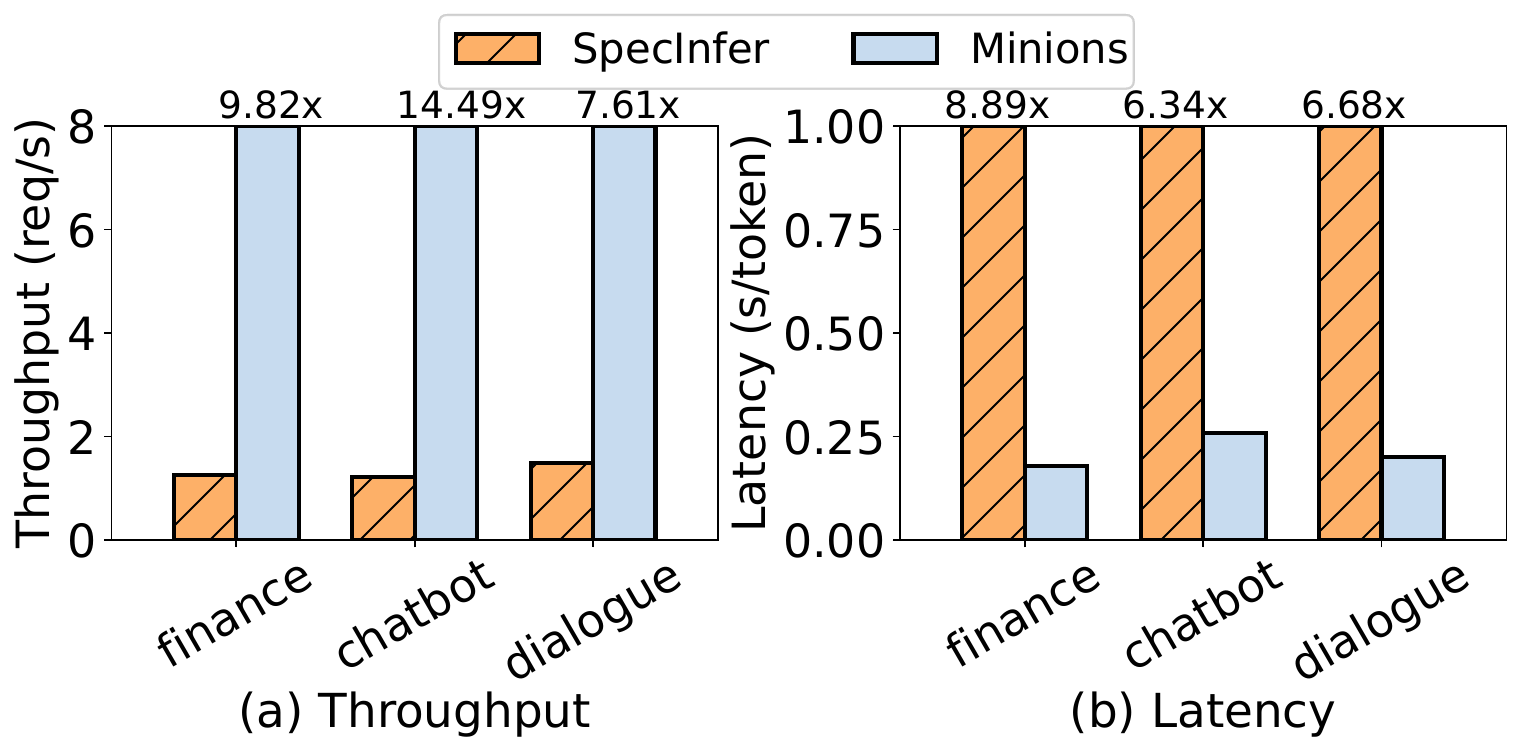}
  \caption{The performance comparison of \textit{Minions} and \textit{SpecInfer}.}
  \label{fig:spec_infer}
\end{figure}

To demonstrate the efficiency of \textit{Minions} compared to existing speculative decoding work, we conduct an experiment with \textit{SpecInfer}~\cite{miao2023specinfer}. The experiment setups are the same as Section~\ref{subsec:setup} with a batch size of 32. As shown in Figure~\ref{fig:spec_infer}, for OPT-13B, \textit{Minions} achieves a speedup of 7.61$\times$ $\sim$ 14.49$\times$ on throughout and 6.34$\times$ $\sim$ 8.89$\times$ on latency. The main reason is that \textit{Minions} efficiently pipelines the execution of SSM and LLM and achieves a high acceptance rate across different datasets via a majority-voted speculator and adaptive speculation length selector. Moreover, the performance of \textit{SpecInfer} is far below what is reported in the paper~\cite{miao2023specinfer}. We suspect the poor performance can be attributed to the problematic operator implementations\footnote{https://github.com/flexflow/FlexFlow/issues/1240}. We have also tried to compare with it on Llama2-70B-chat model. However, it fails to load the attention weights of the model. 

\subsection{Overhead Analysis}

\begin{figure}[htbp]
  \centering
  \includegraphics[width=0.9\columnwidth]{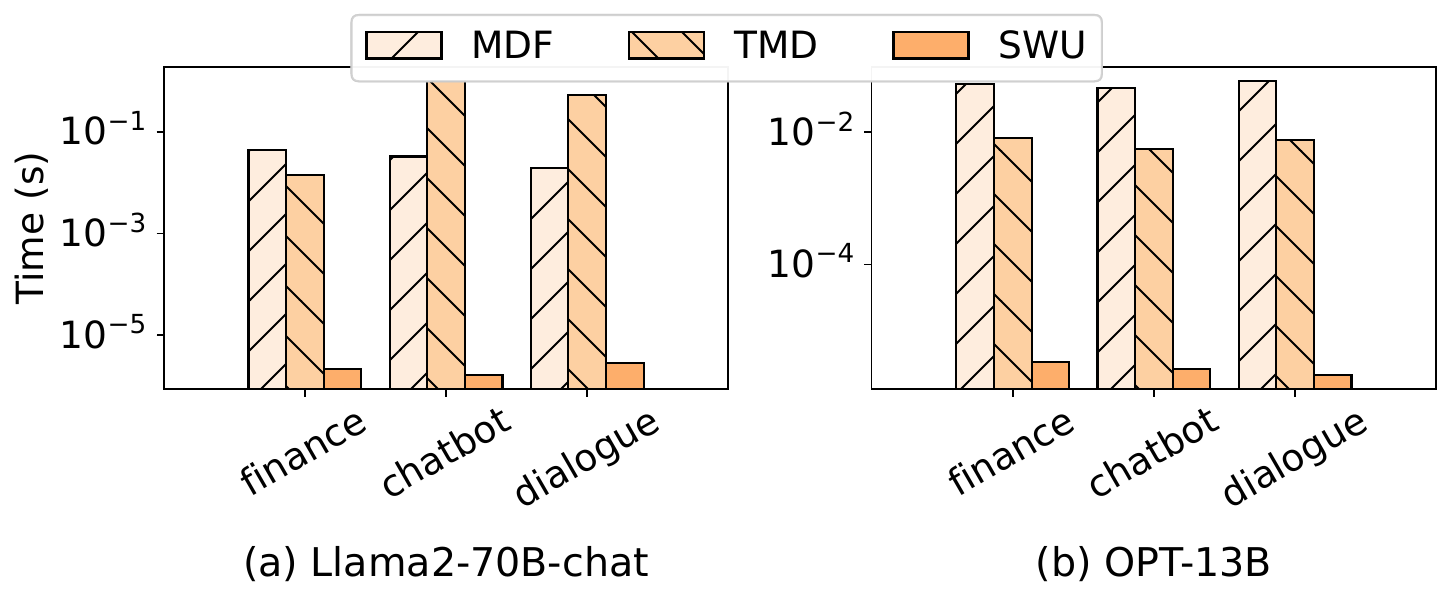}
  \caption{The overhead of \textit{Minions} on two LLMs, including \textit{MDF}, \textit{TMD}, and \textit{SWU}.}
  \label{fig:overhead}
\end{figure}

The overhead of \textit{Minions} can be divided into three parts, including \textit{monitored data fitting} (\textit{MDF}), \textit{tree-based weighted majority decision} (\textit{TMD}), and \textit{SSMs weights update} (\textit{SWU}). \textit{MDF} performs linear fitting on monitored data. \textit{TMD} constructs a tree based on the outputs of SSMs and selects the majority-approved output token by token. \textit{SWU} calculates the acceptance rate for each request and rewards or punishes the weight of the corresponding SSM based on a preset threshold. As illustrated in Figure~\ref{fig:overhead}, the overhead incurred by \textit{MDF}, \textit{TMD} and \textit{SWU} is at most 2.37\% of the total inference time, which is negligible compared to the inference process.

% fine-tune小模型开销？
\section{Related Work}
\label{sec:relatedworks}

Recent work has proposed a series of optimization techniques, which can be categorized into two types: system optimizations and algorithm optimizations.

\textbf{\textit{LLM system optimizations - }}
% vllm, tgi, fasttransformer，trtllm（这个看咱们实验能不能出来），deepspeed-fastgen。我们的方法和框架优化正交，能够adapt到任意框架上。
With the emergence of LLM inference, there have been many efforts to build efficient LLM inference systems~\cite{yu2022orca, kwon2023efficient, tgi, fastgen, tensorrt-llm, sheng2023flexgen, wang2020lightseq}.
% Framework-based system optimizations do not alter the semantics of LLM inference but instead explore the potential of the framework through optimizing scheduling strategies and operator performance. 
Orca~\cite{yu2022orca} proposes iteration-level scheduling, which lowers the scheduling granularity from the request level to the iteration level. Drawing inspiration from virtual memory and paging techniques in operating systems, vLLM~\cite{kwon2023efficient} introduces the PagedAttention to manage KVCache. TGI~\cite{tgi} introduces a new model serialization format safetensor to store the LLM weights. It is faster and more secure compared to other serialization formats. DeepSpeed-FastGen~\cite{fastgen} presents a token composition strategy called Dynamic SplitFuse, which further improves continuous batching and system throughput. TensorRT-LLM~\cite{tensorrt-llm} is built on the inference framework TensorRT and can perform compilation optimizations such as kernel fusion and kernel auto-tuning on computation graphs. The idea behind \textit{Minions} is orthogonal to the above works and can be adapted to these systems.

\textbf{\textit{LLM algorithm optimizations - }}
% 量化 剪枝 spec，前面二者需要对模型进行retrain，产生额外的时间&经济开销，spec仍存在效率低下的问题，我们的xxx
%In addition to system optimizations, there is also a body of work focusing on algorithmic innovations to accelerate LLM inference on GPUs. 
Researchers have proposed various algorithmic innovations to accelerate LLM inference on GPUs~\cite{yao2022zeroquant, lin2023awq, frantar2022optq, jacob2018quantization, dettmers2023qlora, sanh2020movement, michel2019sixteen, fan2019reducing, frantar2023sparsegpt}. Low-bit quantization~\cite{lin2023awq, frantar2022optq, jacob2018quantization} uses fewer bits to represent numerical values of LLM weights and activations. Pruning-based methods~\cite{sanh2020movement, michel2019sixteen, fan2019reducing, frantar2023sparsegpt} reduce the inference cost by setting some weights to zero. While these methods can accelerate model inference, they may also lead to accuracy losses. Other works attempt to improve the process of LLM autoregressive decoding. For instance, blockwise parallel decoding~\cite{stern2018blockwise} and Medusa~\cite{medusa} add new layers on top of the original models to enable parallel inference of multiple tokens at once. However, they need to retrain the model with high economic costs. Compared with these approaches, \textit{Minions} does not require modifying the structure of LLMs or finetuning them. Speculative decoding~\cite{leviathan2023fast, chen2023accelerating} is another approach to accelerate LLM inference. SpecInfer~\cite{miao2023specinfer} combines multiple boost-tuned SSMs to predict the output of the LLM. Although it enhances the efficiency of LLM execution, it concurrently increases the workload of LLM verification, offsetting the performance improvement. In contrast to it, \textit{Minions} leverages a majority-voted mechanism to achieve a well-alignment with LLM without introducing additional verification overhead. 
\section{Conclusion}
\label{sec:conclusion}
In this paper, we propose \textit{Minions}, an LLM inference system with collective and adaptive speculative generation. Specifically, \textit{Minions} proposes a majority-voted mechanism that utilizes multiple SSMs to achieve a well-alignment with LLM. \textit{Minions} also proposes an adaptive mechanism to dynamically adjust the SSM speculation length for better inference performance. Moreover, \textit{Minions} decouple the deployment of LLM and SSM with efficient pipeline execution. The evaluation results demonstrate that \textit{Minions} can achieve higher inference throughput and lower latency compared to existing systems.
%-------------------------------------------------------------------------------
\bibliographystyle{plain}
\bibliography{references}

\begin{thebibliography}{10}

\bibitem{achiam2023gpt}
Josh Achiam, Steven Adler, Sandhini Agarwal, Lama Ahmad, Ilge Akkaya, Florencia~Leoni Aleman, Diogo Almeida, Janko Altenschmidt, Sam Altman, Shyamal Anadkat, et~al.
\newblock Gpt-4 technical report.
\newblock {\em arXiv preprint arXiv:2303.08774}, 2023.

\bibitem{adiwardana2020towards}
Daniel Adiwardana, Minh-Thang Luong, David~R So, Jamie Hall, Noah Fiedel, Romal Thoppilan, Zi~Yang, Apoorv Kulshreshtha, Gaurav Nemade, Yifeng Lu, et~al.
\newblock Towards a human-like open-domain chatbot.
\newblock {\em arXiv preprint arXiv:2001.09977}, 2020.

\bibitem{chen2023accelerating}
Charlie Chen, Sebastian Borgeaud, Geoffrey Irving, Jean-Baptiste Lespiau, Laurent Sifre, and John Jumper.
\newblock Accelerating large language model decoding with speculative sampling.
\newblock {\em arXiv preprint arXiv:2302.01318}, 2023.

\bibitem{chen2021evaluating}
Mark Chen, Jerry Tworek, Heewoo Jun, Qiming Yuan, Henrique Ponde de~Oliveira Pinto, Jared Kaplan, Harri Edwards, Yuri Burda, Nicholas Joseph, Greg Brockman, et~al.
\newblock Evaluating large language models trained on code.
\newblock {\em arXiv preprint arXiv:2107.03374}, 2021.

\bibitem{choi2022serving}
Seungbeom Choi, Sunho Lee, Yeonjae Kim, Jongse Park, Youngjin Kwon, and Jaehyuk Huh.
\newblock Serving heterogeneous machine learning models on multi-gpu servers with spatio-temporal sharing.
\newblock In {\em 2022 USENIX Annual Technical Conference (USENIX ATC 22)}, pages 199--216, 2022.

\bibitem{de2023growing}
Alex de~Vries.
\newblock The growing energy footprint of artificial intelligence.
\newblock {\em Joule}, 7(10):2191--2194, 2023.

\bibitem{dettmers2023qlora}
Tim Dettmers, Artidoro Pagnoni, Ari Holtzman, and Luke Zettlemoyer.
\newblock Qlora: Efficient finetuning of quantized llms.
\newblock {\em arXiv preprint arXiv:2305.14314}, 2023.

\bibitem{fan2019reducing}
Angela Fan, Edouard Grave, and Armand Joulin.
\newblock Reducing transformer depth on demand with structured dropout.
\newblock {\em arXiv preprint arXiv:1909.11556}, 2019.

\bibitem{llama1}
FlexFlow.
\newblock Llama-160m, https://huggingface.co/jackfram/llama-160m, 2023.

\bibitem{frantar2023sparsegpt}
Elias Frantar and Dan Alistarh.
\newblock Sparsegpt: Massive language models can be accurately pruned in one-shot.
\newblock In {\em International Conference on Machine Learning}, pages 10323--10337. PMLR, 2023.

\bibitem{frantar2022optq}
Elias Frantar, Saleh Ashkboos, Torsten Hoefler, and Dan Alistarh.
\newblock Optq: Accurate quantization for generative pre-trained transformers.
\newblock In {\em The Eleventh International Conference on Learning Representations}, 2022.

\bibitem{hsieh2023distilling}
Cheng-Yu Hsieh, Chun-Liang Li, Chih-Kuan Yeh, Hootan Nakhost, Yasuhisa Fujii, Alexander Ratner, Ranjay Krishna, Chen-Yu Lee, and Tomas Pfister.
\newblock Distilling step-by-step! outperforming larger language models with less training data and smaller model sizes.
\newblock {\em arXiv preprint arXiv:2305.02301}, 2023.

\bibitem{huang2020swapadvisor}
Chien-Chin Huang, Gu~Jin, and Jinyang Li.
\newblock Swapadvisor: Pushing deep learning beyond the gpu memory limit via smart swapping.
\newblock In {\em Proceedings of the Twenty-Fifth International Conference on Architectural Support for Programming Languages and Operating Systems}, pages 1341--1355, 2020.

\bibitem{chatbot}
HuggingFace.
\newblock Chatbot, https://huggingface.co/datasets/alespalla/\\ chatbot\_instruction\_prompts, 2023.

\bibitem{finance}
HuggingFace.
\newblock Finance, https://huggingface.co/datasets/gbharti/finance-alpaca, 2023.

\bibitem{tgi}
HuggingFace.
\newblock Text generation inference, https://github.com/huggingface/text-generation-inference, 2023.

\bibitem{jacob2018quantization}
Benoit Jacob, Skirmantas Kligys, Bo~Chen, Menglong Zhu, Matthew Tang, Andrew Howard, Hartwig Adam, and Dmitry Kalenichenko.
\newblock Quantization and training of neural networks for efficient integer-arithmetic-only inference.
\newblock In {\em Proceedings of the IEEE conference on computer vision and pattern recognition}, pages 2704--2713, 2018.

\bibitem{kwon2023efficient}
Woosuk Kwon, Zhuohan Li, Siyuan Zhuang, Ying Sheng, Lianmin Zheng, Cody~Hao Yu, Joseph Gonzalez, Hao Zhang, and Ion Stoica.
\newblock Efficient memory management for large language model serving with pagedattention.
\newblock In {\em Proceedings of the 29th Symposium on Operating Systems Principles}, pages 611--626, 2023.

\bibitem{leviathan2023fast}
Yaniv Leviathan, Matan Kalman, and Yossi Matias.
\newblock Fast inference from transformers via speculative decoding.
\newblock In {\em International Conference on Machine Learning}, pages 19274--19286. PMLR, 2023.

\bibitem{lin2023awq}
Ji~Lin, Jiaming Tang, Haotian Tang, Shang Yang, Xingyu Dang, and Song Han.
\newblock Awq: Activation-aware weight quantization for llm compression and acceleration.
\newblock {\em arXiv preprint arXiv:2306.00978}, 2023.

\bibitem{littlestone1994weighted}
Nick Littlestone and Manfred~K Warmuth.
\newblock The weighted majority algorithm.
\newblock {\em Information and computation}, 108(2):212--261, 1994.

\bibitem{opt-small}
Meta.
\newblock Opt-125m, https://huggingface.co/facebook/opt-125m, 2022.

\bibitem{opt-large}
Meta.
\newblock Opt-13b, https://huggingface.co/facebook/opt-13b, 2022.

\bibitem{llama2}
Meta.
\newblock Llama2-70b-chat, https://huggingface.co/meta-llama/llama-2-70b-chat-hf, 2023.

\bibitem{miao2023specinfer}
Xupeng Miao, Gabriele Oliaro, Zhihao Zhang, Xinhao Cheng, Zeyu Wang, Rae Ying~Yee Wong, Alan Zhu, Lijie Yang, Xiaoxiang Shi, Chunan Shi, Zhuoming Chen, Daiyaan Arfeen, Reyna Abhyankar, and Zhihao Jia.
\newblock Specinfer: Accelerating generative large language model serving with speculative inference and token tree verification, 2023.

\bibitem{michel2019sixteen}
Paul Michel, Omer Levy, and Graham Neubig.
\newblock Are sixteen heads really better than one?
\newblock {\em Advances in neural information processing systems}, 32, 2019.

\bibitem{fastgen}
Microsoft.
\newblock Deepspeed fastgen, https://github.com/microsoft/deepspeed/tree/\\master/blogs/deepspeed-fastgen, 2023.

\bibitem{Chatgpt-prompts}
MohamedRashad.
\newblock Chatgpt-prompts,https://huggingface.co/datasets/mohamedrashad/\\chatgpt-prompts, 2023.

\bibitem{nallapati2016abstractive}
Ramesh Nallapati, Bowen Zhou, Caglar Gulcehre, Bing Xiang, et~al.
\newblock Abstractive text summarization using sequence-to-sequence rnns and beyond.
\newblock {\em arXiv preprint arXiv:1602.06023}, 2016.

\bibitem{mps2012nvidia}
NVIDIA.
\newblock Nvidia: Sharing a gpu between mpi processes: multiple-process service, https://docs.nvidia.com/deploy/mps/index.html, 2012.

\bibitem{tensorrt-llm}
NVIDIA.
\newblock Tensorrt-llm, https://github.com/nvidia/tensorrt-llm, 2023.

\bibitem{openaiprice}
OpenAI.
\newblock Openai pricing,https://openai.com/pricing, 2023.

\bibitem{paulus2017deep}
Romain Paulus, Caiming Xiong, and Richard Socher.
\newblock A deep reinforced model for abstractive summarization.
\newblock {\em arXiv preprint arXiv:1705.04304}, 2017.

\bibitem{peng2020capuchin}
Xuan Peng, Xuanhua Shi, Hulin Dai, Hai Jin, Weiliang Ma, Qian Xiong, Fan Yang, and Xuehai Qian.
\newblock Capuchin: Tensor-based gpu memory management for deep learning.
\newblock In {\em Proceedings of the Twenty-Fifth International Conference on Architectural Support for Programming Languages and Operating Systems}, pages 891--905, 2020.

\bibitem{medusa}
Princeton.
\newblock Medusa, https://sites.google.com/view/medusa-llm, 2023.

\bibitem{rashkin-etal-2019-towards}
Hannah Rashkin, Eric~Michael Smith, Margaret Li, and Y-Lan Boureau.
\newblock Towards empathetic open-domain conversation models: A new benchmark and dataset.
\newblock In {\em Proceedings of the 57th Annual Meeting of the Association for Computational Linguistics}, pages 5370--5381, 2019.

\bibitem{roller2020recipes}
Stephen Roller, Emily Dinan, Naman Goyal, Da~Ju, Mary Williamson, Yinhan Liu, Jing Xu, Myle Ott, Kurt Shuster, Eric~M Smith, et~al.
\newblock Recipes for building an open-domain chatbot.
\newblock {\em arXiv preprint arXiv:2004.13637}, 2020.

\bibitem{sanh2020movement}
Victor Sanh, Thomas Wolf, and Alexander Rush.
\newblock Movement pruning: Adaptive sparsity by fine-tuning.
\newblock {\em Advances in Neural Information Processing Systems}, 33:20378--20389, 2020.

\bibitem{see2017get}
Abigail See, Peter~J Liu, and Christopher~D Manning.
\newblock Get to the point: Summarization with pointer-generator networks.
\newblock {\em arXiv preprint arXiv:1704.04368}, 2017.

\bibitem{sheng2023flexgen}
Ying Sheng, Lianmin Zheng, Binhang Yuan, Zhuohan Li, Max Ryabinin, Beidi Chen, Percy Liang, Christopher R{\'e}, Ion Stoica, and Ce~Zhang.
\newblock Flexgen: High-throughput generative inference of large language models with a single gpu.
\newblock In {\em International Conference on Machine Learning}, pages 31094--31116. PMLR, 2023.

\bibitem{smith2022language}
Ryan Smith, Jason~A Fries, Braden Hancock, and Stephen~H Bach.
\newblock Language models in the loop: Incorporating prompting into weak supervision.
\newblock {\em arXiv preprint arXiv:2205.02318}, 2022.

\bibitem{stern2018blockwise}
Mitchell Stern, Noam Shazeer, and Jakob Uszkoreit.
\newblock Blockwise parallel decoding for deep autoregressive models.
\newblock {\em Advances in Neural Information Processing Systems}, 31, 2018.

\bibitem{sun2022cognn}
Qingxiao Sun, Yi~Liu, Hailong Yang, Ruizhe Zhang, Ming Dun, Mingzhen Li, Xiaoyan Liu, Wencong Xiao, Yong Li, Zhongzhi Luan, et~al.
\newblock Cognn: efficient scheduling for concurrent gnn training on gpus.
\newblock In {\em SC22: International Conference for High Performance Computing, Networking, Storage and Analysis}, pages 1--15. IEEE, 2022.

\bibitem{touvron2023llama}
Hugo Touvron, Louis Martin, Kevin Stone, Peter Albert, Amjad Almahairi, Yasmine Babaei, Nikolay Bashlykov, Soumya Batra, Prajjwal Bhargava, Shruti Bhosale, et~al.
\newblock Llama 2: Open foundation and fine-tuned chat models.
\newblock {\em arXiv preprint arXiv:2307.09288}, 2023.

\bibitem{vaswani2017attention}
Ashish Vaswani, Noam Shazeer, Niki Parmar, Jakob Uszkoreit, Llion Jones, Aidan~N Gomez, {\L}ukasz Kaiser, and Illia Polosukhin.
\newblock Attention is all you need.
\newblock {\em Advances in neural information processing systems}, 30, 2017.

\bibitem{wang2020lightseq}
Xiaohui Wang, Ying Xiong, Yang Wei, Mingxuan Wang, and Lei Li.
\newblock Lightseq: A high performance inference library for transformers.
\newblock {\em arXiv preprint arXiv:2010.13887}, 2020.

\bibitem{xu2015show}
Kelvin Xu, Jimmy Ba, Ryan Kiros, Kyunghyun Cho, Aaron Courville, Ruslan Salakhudinov, Rich Zemel, and Yoshua Bengio.
\newblock Show, attend and tell: Neural image caption generation with visual attention.
\newblock In {\em International conference on machine learning}, pages 2048--2057. PMLR, 2015.

\bibitem{yang2016review}
Zhilin Yang, Ye~Yuan, Yuexin Wu, William~W Cohen, and Russ~R Salakhutdinov.
\newblock Review networks for caption generation.
\newblock {\em Advances in neural information processing systems}, 29, 2016.

\bibitem{yao2022zeroquant}
Zhewei Yao, Reza Yazdani~Aminabadi, Minjia Zhang, Xiaoxia Wu, Conglong Li, and Yuxiong He.
\newblock Zeroquant: Efficient and affordable post-training quantization for large-scale transformers.
\newblock {\em Advances in Neural Information Processing Systems}, 35:27168--27183, 2022.

\bibitem{yu2022orca}
Gyeong-In Yu, Joo~Seong Jeong, Geon-Woo Kim, Soojeong Kim, and Byung-Gon Chun.
\newblock Orca: A distributed serving system for $\{$Transformer-Based$\}$ generative models.
\newblock In {\em 16th USENIX Symposium on Operating Systems Design and Implementation (OSDI 22)}, pages 521--538, 2022.

\bibitem{zhang2022opt}
Susan Zhang, Stephen Roller, Naman Goyal, Mikel Artetxe, Moya Chen, Shuohui Chen, Christopher Dewan, Mona Diab, Xian Li, Xi~Victoria Lin, et~al.
\newblock Opt: Open pre-trained transformer language models.
\newblock {\em arXiv preprint arXiv:2205.01068}, 2022.

\end{thebibliography}

%%%%%%%%%%%%%%%%%%%%%%%%%%%%%%%%%%%%%%%%%%%%%%%%%%%%%%%%%%%%%%%%%%%%%%%%%%%%%%%%
\end{document}